\documentclass[draft]{agujournal2019}
\usepackage{url} 
\usepackage{lineno}
\usepackage[inline]{trackchanges} 
\usepackage{soul}
\usepackage{siunitx} 
\usepackage{amsmath} 
\usepackage{bbm} 
\usepackage{amsfonts} 
\usepackage{booktabs} 

\draftfalse

\journalname{Journal of Advances in Modeling Earth Systems (JAMES)}

\begin{document}

\title{Stress-Testing Dynamical and Generative Downscaling Using Subseasonal Extreme Precipitation Forecasts}

\authors{Mauricio Lima\affil{1,2}, Marika Koukoula\affil{3}, Romain Pilon\affil{3}, Monika Feldmann\affil{4,5}, Erwan Koch\affil{1,3}, Daniela I.V. Domeisen\affil{3}, Tom Beucler\affil{1,3}}

\affiliation{1}{Expertise Center for Climate Extremes, University of Lausanne, Switzerland}
\affiliation{2}{Max Planck Institute for Biogeochemistry, Germany}
\affiliation{3}{Faculty of Geosciences and Environment, University of Lausanne, Switzerland}
\affiliation{4}{Institute of Geography - Oeschger Centre for Climate Change Research, University of Bern, Switzerland}
\affiliation{5}{Institute for Atmospheric and Climate Science, ETH Zurich, Zurich, Switzerland}

\correspondingauthor{Mauricio Lima}{mlima@bgc-jena.mpg.de}


\begin{keypoints}
\item We assess the WRF model and a generative diffusion model for downscaling two high-impact precipitation events for lead times up to 3 weeks.

\item Both downscaling methods improve the probabilistic calibration and spatial distribution of the baseline forecasts provided by ECMWF.

\item While WRF excels in a multicell event, the diffusion model is more consistent across the two events and across performance metrics.
\end{keypoints}

\begin{abstract}
Coarse spatial resolution limits the ability of subseasonal prediction models to resolve extreme precipitation. Downscaling with either dynamical or deep generative models can overcome this issue, but the comparative performance of these models for extremes across different atmospheric regimes remains poorly understood. In this work, we evaluate the Weather Research and Forecasting (WRF) model against a diffusion-based generative model by downscaling two physically distinct, extreme precipitation events up to lead times of 3 weeks. For a fair comparison with WRF, which can downscale boundary conditions from different driving models without model-specific training, the diffusion model is trained in an unpaired fashion. Both approaches improve upon the raw European Centre for Medium-Range Weather Forecasts forecasts, in comparison to fused rain gauge-radar observations in Switzerland (CombiPrecip), but exhibit regime-dependent strengths. WRF achieves the highest probabilistic skill for a multicell, non-stationary event. Conversely, the diffusion model is more consistent across different performance metrics for the two events, outperforming WRF in a more stationary supercell event. These results demonstrate that explicit dynamical modeling can add value for specific precipitation events for subseasonal lead times, and that generative downscaling adds value more broadly in different situations.
\end{abstract}

\section*{Plain Language Summary}
In the context of forecasting severe storms several weeks ahead, we compare state-of-the-art methods that take completely different approaches: dynamical (physics-based) versus generative (AI-based) downscaling. Downscaling, which consists of improving the location and spatial resolution of forecasts (``zooming in''), has great potential for subseasonal time scales (here up to three weeks in advance), since subseasonal physics-based models are too blurry to represent severe storms well. We analyze two case studies of storms associated with extreme precipitation. Dynamical downscaling excels in one specific event, while generative downscaling is more consistent across the two events. Both methods improve performance metrics compared with directly using the raw coarse forecasts.

\section{Introduction}

Weather forecasting models typically predict key atmospheric variables up to two weeks into the future and are essential to anticipate extreme events \cite{bauer2015quiet}. Climate models simulate longer time scales, and are rather used to evaluate the statistics of changes in distributions \cite{ogorman2009scaling}. In between these two scales, subseasonal forecasts, covering the forecasting range from 2 weeks to 2 months, have progressed greatly in the past decade, but remaining challenges still need to be addressed \cite{vitart2025wwrp}: improving the accuracy and resolution of these forecasts is vital for a wide range of applications, including agriculture and water resource management \cite{merryfield2020current}. Even though the two-week window is known to be a difficult barrier to surpass in terms of timing and spatial localization of events \cite{bauer2015quiet, domeisen2018predictable}, these forecasts could be used to anticipate and manage extreme events \cite{domeisen2022advances, white2022advances}, some of which have shown an increasing trend under climate change \cite{ipcc2021chapter11}. To tackle this problem, reducing the bias generated by the unresolved processes---most prominently convection---of such models is necessary. Climate models have historically overcome similar tasks using statistical downscaling techniques, which translate large-scale model outputs into local-scale information that better represents regional features \cite{fowler2007linking}. Just like climate models, subseasonal prediction systems use global coverage due to the global-scale remote processes that can impact local predictability \cite{stan2017review,vitart2019introduction}, but the need for global coverage comes at the expense of local resolution.  While downscaling is a well-established technique for climate projections, applying these methods to the subseasonal timescale remains under-explored.

A primary physical motivation for subseasonal downscaling is the explicit representation of deep convection \cite{castro2020convective}. At subseasonal time scales, operational global models capture large-scale synoptic dynamics explicitly, but lack the spatial resolution to explicitly represent mesoscale convective phenomena. The development of severe thunderstorms requires thermodynamic environments characterized by high atmospheric instability and deep-layer wind shear \cite{houze2014cumulonimbus, feldmann2023types, feldmann2024modeling}. However, the coarser resolution of subseasonal models poorly resolves local instabilities and shear and relies on parameterized convection that does not represent local extreme precipitation and its consequences accurately. Extreme precipitation is a direct result of updrafts that are not resolved in these models and that are involved in the formation of severe thunderstorms, and serves as a fundamental variable to evaluate the downscaling performance. To bridge this gap, one can typically choose between dynamical and statistical downscaling.

Dynamical downscaling is a technique that consists of nesting one or multiple high-resolution physics-based models within coarser-resolution data, which usually come from a coarser, but geographically broader model \cite{giorgi2015regional, xu2018dynamical}. The coarser-resolution data are responsible for providing large-scale information (e.g., synoptic-scale dynamics), which serves as the boundary condition for the high-resolution model, allowing it to capture small-scale features and processes that are not well resolved at larger scales (e.g., convection). Particularly in the context of extreme precipitation, strong links have been identified between large-scale patterns and severe convective outbreaks \cite{feldmann2025pan}. More generally, this approach allows for the simulation of complex weather patterns at a higher level of detail, improving the accuracy and reliability of predictions without losing the physical consistency, i.e., having conservation of mass, momentum and energy also at the high-resolution domain. Additionally, dynamical downscaling allows for a three-dimensional representation of the atmosphere, and at fine grid resolutions, explicitly resolves convection. Notably, the Weather Research and Forecasting \cite<WRF;>{powers2017wrf} model is widely adopted for downscaling and modeling extreme precipitation events \cite{walton2020wrf, merino2022wrf, draeger2024wrf, rahimi2024wrf} and is also our choice for dynamical downscaling in this work. A general downside of dynamical models is their computational demand.

Statistical downscaling, on the other hand, employs statistical relationships between high- and low-resolution variables and is associated with considerably faster computations. Historically, one of the most common statistical methods for downscaling has been bias correction and statistical disaggregation (BCSD), which first interpolates coarse model outputs to match the high resolution from observations and then applies bias correction---typically using quantile mapping \cite{panofsky1958some, maraun2013bias}---to adjust the statistical distribution of the modeled variables. However, these techniques may not capture all the nuances of physical interactions as effectively as dynamical models. Recently, advances in machine learning have shown promising new approaches for downscaling \cite{rampal2024enhancing}. In particular, generative models are well-suited to tackling extreme events, since these models naturally provide ensembles representing a probability distribution. For example, generative adversarial networks \cite<GANs;>{goodfellow2020generative}, have been used for downscaling in, e.g., \citeA{miralles2022downscaling} and \citeA{price2022increasing}. Denoising diffusion probabilistic models \cite{sohl-dickstein2015deep,ho2020denoising}, or ``diffusion models'' for short, have been used for downscaling in, e.g., \citeA{mardani2023residual} and \citeA{bischoff2024unpaired}, and are starting to be competitive with traditional downscaling methods when employed to subseasonal prediction systems \cite{pyrina2026joint}. We decided to adopt diffusion models in our work due to their success in downscaling when compared to other methods \cite{wan2024debias}.

By focusing on extreme precipitation, one of the least predictable weather phenomena at subseasonal timescales \cite{domeisen2022advances}, we put the current generation of downscaling models to a genuine test. The novelty of this work lies in the contrast of state-of-the-art dynamical and generative approaches on two severe convective events with different storm morphology: multicell thunderstorms and supercells with upscale growth. Moreover, despite recent work using diffusion models to downscale wind at the subseasonal scale \cite{springenberg2026diffscale}, downscaling of subseasonal forecasts is still under-explored, in particular for the challenging task of predicting extreme precipitation.

The remainder of the paper is organized as follows. Section \ref{sec:data} (Data) describes the data used as baseline, as input to the downscaling models, and for validating their performance. Section \ref{sec:model-evaluation} (Model evaluation) outlines the metrics used to quantify and compare model performance. Section \ref{sec:dynamical-downscaling} (Dynamical downscaling) details the configuration of the dynamical model, while Section \ref{sec:generative-downscaling} (Generative downscaling) presents the generative model and its underlying framework. Section \ref{sec:results} (Results) reports the outcomes obtained for each of the selected metrics, Section~\ref{sec:discussion} (Discussion) contrasts the main differences between the two approaches, and Section~\ref{sec:conclusion} summarizes the main findings and directions for future research.

\section{Data}
\label{sec:data}

Here we provide a physical description of the chosen weather events, then present the datasets used to perform the downscaling tasks.

\subsection{Meteorological Description}
\label{sec:physical-description}

The chosen extreme events are the thunderstorms that struck Switzerland on 11--12 June 2018 and 28--29 June 2021. While both rank among the most intense precipitation events in the dataset of \citeA{feldmann2022radar}, they were characterized by different atmospheric dynamics and convective modes. The 2018 event, which set the Swiss national 10-minute precipitation record, was embedded in a broader sequence of severe convective storms affecting Europe throughout May and June 2018 \cite{mohr2020role,liernur2025small}. A blocking situation over Scandinavia \cite[Figure 5]{mohr2020role} advected moist and warm air into western Europe, where weak horizontal gradients in both pressure and geopotential heights inhibited large-scale circulation, favoring the persistence of these air masses over the region. This also resulted in low wind speeds throughout the troposphere, causing thunderstorms to remain nearly stationary, and thereby producing torrential rain accumulation. Ultimately, the presence of a broad upper-level trough over southwestern France, containing several embedded short-wave disturbances on 11 June, provided the final trigger for deep convection. The resulting convective system was characterized by a multicellular configuration, in which individual cells could trigger new convection through cold-pool interactions. Although the event was embedded in a favorable synoptic-scale environment, its extreme precipitation was thus associated with the organization and successive triggering of convective cells. In contrast, the 2021 event was characterized by a different convective mode, with a supercell developing into a bow echo and producing severe hail impacts across Switzerland with return periods exceeding the 100-years range \cite{kopp2023summer}. During this event, high convective available potential energy (CAPE), integrated moisture content, and bulk shear created an environment highly conducive to the development and organization of the supercell. Synoptic-scale forcing also contributed to the event through the presence of a positive potential vorticity anomaly over western France and a downstream ridge that promoted the advection of warm and moist air towards Switzerland. The 2021 forecast therefore traces the convective system as it propagated across Switzerland.

\subsection{Datasets}
\label{sec:datasets}
\begin{figure}
 \centering
 \includegraphics[width=\linewidth]{
 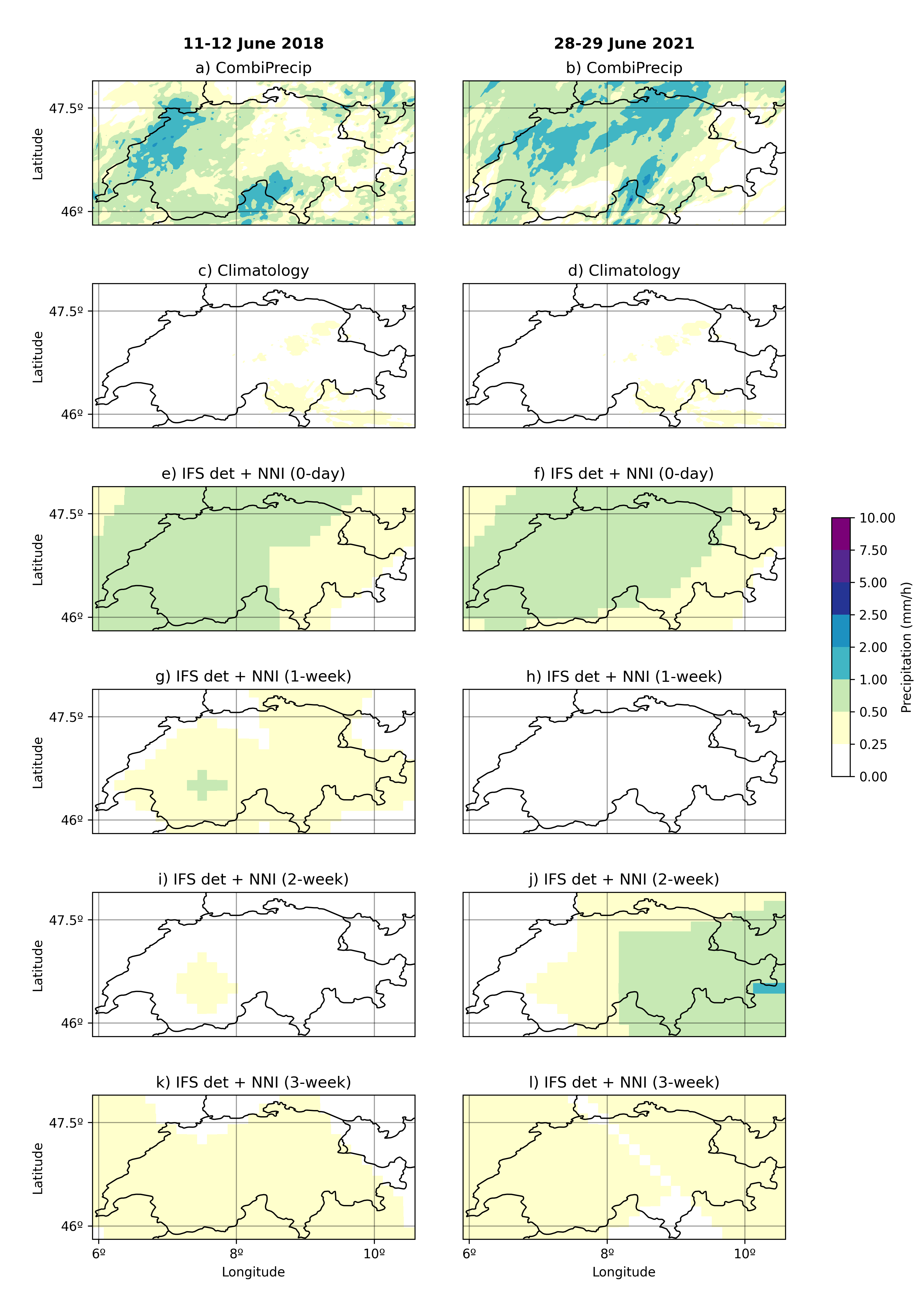}
 \caption{(a-l) Mean aggregated precipitation for the 2018 (left column) and 2021 (right column) extreme precipitation events. Rows from top to bottom display (a-b) the fused gauge-radar CombiPrecip ground truth, (c-d) the climatology, which is the same for both events, and the deterministic IFS forecast with a nearest neighbor interpolation (NNI) for the lead times of (e-f) 0-day, (g-h) 1-week, (i-j) 2-weeks, and (k-l) 3-weeks. A nearest neighbor interpolation is performed to match the resolution in CombiPrecip, which is done for evaluation purposes.}
 \label{fig:agg-maps-raw}
\end{figure}

Figure~\ref{fig:agg-maps-raw} illustrates the broad objective of this study: bridging the gap between the low-resolution forecasts produced by the subseasonal Integrated Forecasting System (IFS) products from the European Centre for Medium-Range Weather Forecasts (ECMWF), and the high-resolution observational dataset CombiPrecip over the climatology of the period for Switzerland. The two approaches in question---dynamical and generative---use the information available from the forecasts to enhance the original skill and will be compared to CombiPrecip. The figure contrasts the two events: in 2018 (on the left) and in 2021 (on the right). The CombiPrecip aggregates (Figure~\ref{fig:agg-maps-raw}a-b) show the most affected regions in each of the events. The climatology is the same for both years (Figure~\ref{fig:agg-maps-raw}c-d). The 0-day forecasts (Figure~\ref{fig:agg-maps-raw}e-f) display a greater aggregate of precipitation in comparison to the more dispersed forecasts at the greater lead times (Figure~\ref{fig:agg-maps-raw}g-l). Next, we describe each data product in detail.

We choose CombiPrecip \cite{sideris2013real} as the high-resolution dataset and observational reference (``ground truth'') for the comparison of our models, a state-of-the-art dataset for precipitation in Switzerland. It provides detailed and accurate gridded precipitation estimates at the ground through the combination of radar and automated gauge data. The dataset is used at a spatial resolution of $1$~\unit{km} and at a time resolution of $1$~\unit{h}, but aggregated to 6-hourly resolution to match the time resolution offered by the subseasonal forecasts described next. The dataset is provided using the Swiss coordinate system (CH1903+ / LV95), which we transform to latitude and longitude for comparison with the other outputs, while keeping the same resolution. To match a regular grid, we re-grid the domain using a two-dimensional linear interpolation. We select longitudes between $5.9^{\circ}$ and $10.6^{\circ}$ and latitudes between $45.8^{\circ}$ and $47.9^{\circ}$. This spatial domain essentially depicts the geographical location of Switzerland, which is shown in Figure~\ref{fig:agg-maps-raw}a, depicting the observed precipitation on 11-12 June 2018. As a quality control, we trim values of precipitation between 0.0~mm/h and 0.1~mm/h to 0.0~mm/h, given the uncertainty of radar measurements \cite{gabella2018} for metric computations in the evaluation phase.

The subseasonal forecasts of precipitation to be downscaled come from the operational stream of the subseasonal data base from ECMWF; see \citeA{vitart2017subseasonal}. More specifically, model version cy43r3 is used for the event in 2018 and model version cy47r2 for the event in 2021. We use the model output from the control run, providing one deterministic forecast (IFS det) and the perturbed runs, providing 50 ensemble members (IFS ens). In our study, the forecasts are verified in a 48~h-window starting at 00:00~UTC of the first day of each event: 11 June 2018 and 28 June 2021. We consider 4 initialization dates: 0-day, 1-, 2-, and 3-weeks in advance of the events, also taking 00:00~UTC of the first event as the reference. This choice balances practicality and relevance: the 3-week horizon extends beyond the typical 2-week limit of weather forecasts, stepping inside the subseasonal timescale. At the same time, convective summer events are known to have poor predictability, so we also keep the shorter lead time of 1-week in our experiments, and we keep the 2-week lead time as an intermediate window. We additionally provide the 0-day forecast---initialization at 00:00~UTC of the start of the 48~h verification period---for comparison with CombiPrecip. The four lead times are shown in Figure~\ref{fig:agg-maps-raw}e-l. The forecast output has a time resolution of $6$~\unit{h} for the surface variables and daily resolution for the pressure level variables. Precipitation is part of the surface variables, hence the time resolution for the (spatial) downscaling task in this article is 6-hourly. The forecasts are available at a global resolution of $0.15$\unit{\degree}, which corresponds to approximately $11$~\unit{km} in latitude and $16$~\unit{km} in longitude in Switzerland. 

Different than the dynamical model, the generative model (diffusion) does not receive input from larger-area boundary conditions nor from variables other than precipitation (i.e., we perform a super-resolution task). We detail the domain and variables of the dynamical model in Section~\ref{sec:dynamical-downscaling}. We define the training set of the model as the CombiPrecip data for the summer months (June, July and August) between 2018 and 2023, removing the target test dates for the two events described above. For means of comparison, we also display the climatology, taken as the empirical mean of the training set of the generative downscaling model for each pixel, which is shown in Figure~\ref{fig:agg-maps-raw}c-d. We will also use this climatology as the baseline forecast to evaluate the skill score in Section~\ref{sec:results}.

\section{Model evaluation}
\label{sec:model-evaluation}
To evaluate the models, we choose metrics that characterize the quality of the models under different angles, looking first at a proper metric for ensembles \cite{gneiting2007strictly}, then at interpretability from the meteorological point of view. One important feature to consider is the size of the ensemble produced by each of the models. The models considered here generate ensembles of different sizes, which is a consequence of their computational cost. Indeed, because generative models are less computationally expensive as compared to dynamical models, they can produce a larger number of ensemble members.

Before delving into each metric, we highlight that most metrics (all but the spread skill ratio, explained in Section~\ref{sec:spread-skill-ratio}) are taken as the skill score (SS) with respect to the climatology (which is explained in detail in Section~\ref{sec:data}). This is done to contextualize performances, similarly to \citeA{li2021improvements}. Mathematically, for a given error or distance metric $m$ (where the perfect value would be 0), the skill score is generally given by
\begin{align}
 \text{SS}_m = 1 - \frac{m_{\text{forecast}}}{m_{\text{climatology}}}.
\end{align}
Using this formulation, a perfect forecast has a skill score of 1 and negative values indicate performance worse than climatology.

\subsection{Continuous Ranked Probability Score}
\label{sec:crps}
The continuous ranked probability score (CRPS) is a proper scoring metric \cite{gneiting2007strictly}, meaning it maximizes the expected score when the ensemble distribution accurately represents the true distribution.

Let $x_{i,t,n}$ represent the forecast of the $n$-th ensemble member at grid point $i$ and time $t$. Let $y_{i,t}$ be the corresponding observed values for the same grid point and time. The empirical fair CRPS is calculated as
\begin{align}
\label{eq:empirical-crps}
\text{CRPS} = \left\langle \frac{1}{N} \sum_{n=1}^N |x_{i,t,n} - y_{i,t}| - \frac{1}{2N(N-1)} \sum_{n=1}^N \sum_{n'=1}^N |x_{i,t,n} - x_{i,t,n'}| \right\rangle_{i,t},
\end{align}
as proposed in \citeA{ferro2014fair}, where $\langle\cdot\rangle_{i,t}$ denotes the mean over grid points and time. The first term represents the mean absolute error between the ensemble members and the observation, while the second term measures the internal ensemble spread. By using the denominator $N(N-1)$, the score becomes unbiased, which, as mentioned before, is necessary to guarantee a fair comparison between products with different ensemble member sizes. For a deterministic forecast ($N=1$), the CRPS is equivalent to the mean absolute error, since we can simply ignore the right term on the right hand side in \eqref{eq:empirical-crps}.

In theory, the CRPS already suffices to evaluate the overall probabilistic skill of the models, as this is a proper scoring metric. However, we add several other metrics to better be able to interpret the specific physical and spatial differences between the modeling approaches.

\subsection{Member-Averaged Fraction Skill Score}
\label{sec:member-averaged-fss}

The deterministic fraction skill score \cite<FSS,>[]{roberts2008scale} is adopted in the weather forecasting community when analyzing forecasts that may suffer from displacement errors. For instance, a model may not reproduce the exact precipitation at one specific grid point perfectly, but still capture the general behavior of precipitation in the neighborhood of this grid point, in which case it should have a good evaluation score. To analyze the spatial quality of the models, we choose the member-averaged fraction skill score (avFSS), which has a constant behavior for different ensemble sizes~\cite{necker2024fractions}.

The avFSS can be formulated as
\begin{align}
 \label{eq:avfss}
 \text{avFSS}(q,J) = 
 1 - 
 \frac{1}{N}\sum_{n=1}^{N}
 \frac{\sum_{w}[P^f_n(q,w(J))-P^o(q,w(J))]^2}{\sum_{w}P^f_n(q,w(J))^2+\sum_{w}P^o(q,w(J))^2},
\end{align}
where $P^f_n(q,w(J))$ represents the fraction of precipitation values above a threshold value $q$ for a window $w$ with $J$ grid points for the forecast member $n$ and $P^o(q,w(J))$ represents a similar fraction of precipitation, but for the observations. The numerator in the formula is the fraction Brier Score \cite{glenn1950verification} for one ensemble member, and the denominator is a measure to obtain a skill score from it. The FSS, and consequently the avFSS, is defined in the $[0,1]$ range. A forecast with no skill will have an FSS of 0 and a perfect forecast will have an FSS of 1.

In (\ref{eq:avfss}), the sum $\sum_{w}$ is computed over all windows available with size $J$ in both forecast and observations. Thereby, we highlight that the avFSS has only two free parameters: the neighborhood size $J$ and the threshold $q$.

The neighborhood size $J$ defines the number of grid points per neighborhood (of each grid point $i$). In this study, we use ensembles at the kilometer scale resolution, even though the original IFS forecasts to be downscaled have approximately $15$~\unit{km} resolution. This motivates us to use $J=15\times 15$ as the neighborhood size.

The threshold $q$ defines the event whose spatial occurrence is being evaluated. Naturally, more extreme precipitation events  imply higher thresholds and should have a lower skill score. As our main goal is to better predict extreme events, we  explore higher values of threshold, although precipitation extremes are difficult to predict and the score should decrease significantly for too high values of $q$.

\subsection{Wasserstein Distance}
\label{sec:wasserstein-distance}
To infer the distribution quality of the generated forecasts, we use the Wasserstein distance between the cumulative distribution functions (CDFs) over the entire set, i.e., aggregating ensemble members, time and space dimensions for each lead time. The distance can be written as 
\begin{align}
 W_1 = \int
 \left |
 F^f(x) - F^o(x)
 \right | \text{d}x,
\end{align}
where the one-dimensional CDF, denoted by $F$, should be integrated over the mathematical domain of definition. In practice, the integral is only computed from the minimum value of precipitation ($0~\unit{mm/h}$) to the maximum value (around $25~\unit{mm/h}$, which is significantly smaller than peak values registered during the events as we here work with the 6-hourly average). The index $1$ of $W_1$ denotes the use of the $l_1$ distance between the two curves. It is worth noting that W1 is a measure for the entire distribution, even though we are here mostly interested in the right tail (extreme values).

\subsection{Spectral Distance}
\label{sec:spectral-distance}
The spectral evaluation targets the skill of each model to generate the adequate amount of information (variance) at the correct scales. Regarding the quality of the spatial scales reproduced by the downscaling methods, one can expect to look at the power spectral density (PSD). Just like before, the PSD distance ($\text{PSDd}$) for a deterministic forecast can then be written as 
\begin{align}
 \text{PSDd} = \iint
 \left |
 \text{PSD}^f(k_x, k_y) - \text{PSD}^o(k_x, k_y)
 \right | \text{d}k_x \text{d}k_y,
\end{align}
where the two-dimensional PSD is taken over the spatial dimensions and averaged over time, and the integral is computed in the wavenumber domain defined by the forecast or observation. $\text{PSD}^f$ and $\text{PSD}^o$ stand for the PSD of the forecast and the observations respectively. We take the $l_1$ norm to align with the norm taken in W1. To extend this metric for ensembles, we again proceed like for the avFSS, leading us to define the member-averaged PSDd metric
\begin{align}
 \text{avPSDd} = \frac{1}{N}\sum_{n=1}^{N}\text{PSDd}.
\end{align}

\subsection{Spread-Skill Ratio}
\label{sec:spread-skill-ratio}

To evaluate the probabilistic calibration of the generated ensembles, we compute the Spread-skill ratio (SSR) which can be formulated as 

\begin{align}
\text{SSR} = \sqrt{\frac{N+1}{N-1}} \dfrac{\sqrt{\left\langle\frac{1}{N} \sum_{n=1}^N (x_{i,t,n} - \langle x_{i,t}\rangle_n)^2 \right\rangle_{i,t}}}{\sqrt{\left\langle(\langle x_{i,t}\rangle_n - y_{i,t})^2\right\rangle_{i,t}}},
\end{align}

where $\langle x_{i,t}\rangle_n$ denotes the mean over the ensemble. The left term on the right hand side is a correction term to make the dispersion metric invariant to the ensemble size, as motivated by \citeA{roberts2025unbiased}. 

An SSR of $1$ indicates perfect dispersion, where the internal ensemble variance accurately reflects the true forecast error. Under-dispersive forecasts have $\text{SSR} < 1$ and over-dispersive forecasts have $\text{SSR} > 1$. The reason why the SSR is not used as a skill score for the climatology (or any deterministic forecast) is because it is not defined for deterministic products.

\section{Downscaling Models}

\subsection{Dynamical Downscaling}
\label{sec:dynamical-downscaling}

We use version 4.4 of WRF \cite{skamarock2021wrf}, which was developed at the National Center for Atmospheric Research (NCAR). Simulations are performed with the Advanced Research WRF dynamical core, which solves the fully compressible non-hydrostatic equations on a terrain-following vertical coordinate using a time-split integration scheme. WRF makes use of multiple variables, which we list in Table~\ref{tab:inputs} spreading over a wide domain, which we describe next.

\begin{figure}
 \centering
 \includegraphics[width=0.5\textwidth]{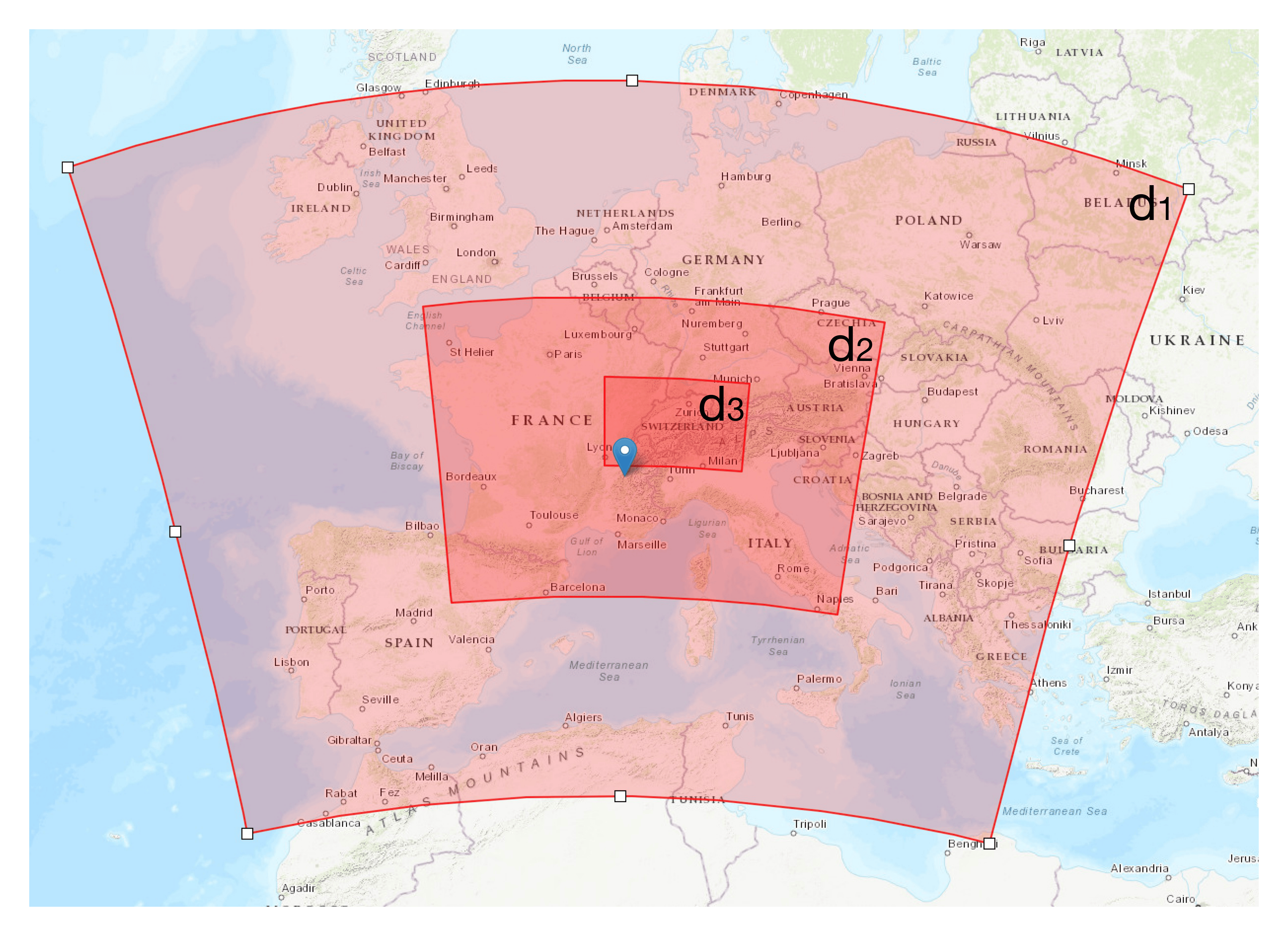}
 \caption{Domains $d_1$ ($9$~\unit{km} $\times$ $9$~\unit{km}), $d_2$ ($3$~\unit{km} $\times$ $3$~\unit{km}) and $d_3$ ($1$~\unit{km} $\times$ $1$~\unit{km}) in increasing order of resolution and decreasing order of size used to run the WRF experiments. To choose and visualize these domains we used \protect\url{https://jiririchter.github.io/WRFDomainWizard/}. The blue dot in the map marks the center of $d_1$.}
 \label{figs:wrf-domains}
\end{figure}

We adopt three nested domains following a ratio of $1/3$ between the resolution of inner and outer domains. We set a parent domain with a resolution of $9$~\unit{km} $\times$ $9$~\unit{km} ($d_1$), a first inner domain with a resolution of $3$~\unit{km} $\times$ $3$~\unit{km} ($d_2$) and a second inner domain with a resolution of $1$~\unit{km} $\times$ $1$~\unit{km} ($d_3$), which are displayed in Figure~\ref{figs:wrf-domains}. To choose these domains on the map, we avoid crossing steep topography lying exactly on the border of a domain. We adopt the one-way nesting approach, i.e., the inner domains use the boundary conditions of the outer domains and the outer domains are not influenced by the inner domains. We use $35$ vertical levels on a terrain-following, vertically stretched coordinate, with the lowest layer at 20 m above the surface. Layer thickness increases with a surface stretch factor of $1.2$ near the ground and $1.1$ in the upper atmosphere, reaching a maximum thickness of $1000$~\unit{m}. The model top is set at 50 hPa.

We set a spin-up time of $12$~\unit{h} to allow the model to reach a stable state \cite<similarly to, e.g.,>{chu2018evaluation, gomez2018new}. Furthermore, we update the boundary conditions of the coarser outer domain ($d_1$) with the IFS data using the same resolution as the pressure level data, i.e., 24-hourly. We use a time step of $9$~\unit{s} for the domain $d_1$ and $4$~\unit{s} for both domains $d_2$ and $d_3$, small enough to guarantee stable integration (e.g., satisfy the Courant number).

Next, we describe the sub-km processes parameterizations. For microphysical processes, we use the ``double-moment" Thompson Scheme \cite{thompson2008explicit} across all three domains. This scheme offers a representation of cloud microphysics, including ice, snow, and graupel, and provides a balance between computational efficiency and physical realism.
Owing to the resolution of the parent domain, cumulus convection is parametrized \cite{kyle1976fitting, wang2020updraft}, while the inner domains do not require parametrization of convection.

For the planetary boundary layer (PBL), we use the Yonsei University Scheme \cite{hong2006diffusion} for all three domains. For the surface layer scheme, the Revised MM5 Scheme \cite{jimenez2012revised} is used for all three domains. This scheme is responsible for parameterizing the fluxes of heat, moisture, and momentum between the surface and the atmosphere, impacting near-surface weather conditions.

For the land surface, we utilize the Noah–MP Land Surface Model \cite{niu2011community, yang2011community} across all domains. This model includes features such as multiple layers of soil and vegetation. For long- and shortwave radiation, we employ the RRTMG Schemes \cite{iacono2008radiative} for all three domains.

Finally, given Switzerland's large number of lakes and their importance for convection \cite{feldmann2024modeling}, we enable explicit lake interactions over all three domains. This inclusion allows for a more accurate representation of lake effects on local weather patterns, such as humidity and temperature regulation, which can influence regional climate and weather forecasts.

\begin{figure}
 \centering
 \includegraphics[width=\textwidth]{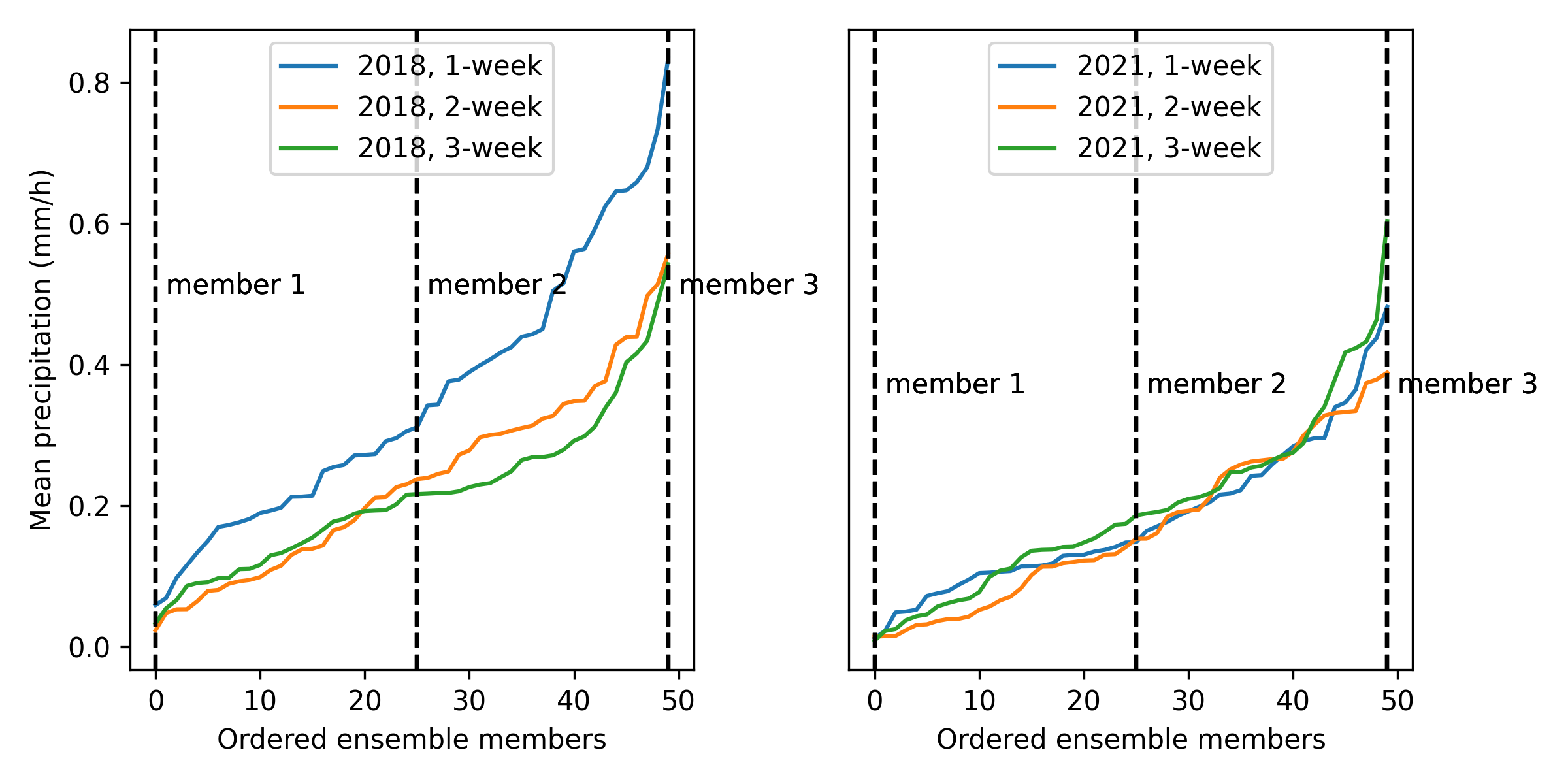}
 \caption{Ensemble members from the IFS perturbed forecast for the event on 11-12 June 2018 (left panel) and for the event on 28-29 June 2021 (right panel), ordered by mean precipitation. The 3 chosen members to be used as boundary conditions are chosen to maximize the spread and are depicted as black dashed lines.}
 \label{figs:ordered-spread}
\end{figure}

In order to generate different ensemble members with WRF, we use boundary conditions coming from different IFS ensemble members. Because dynamical downscaling with WRF is computationally demanding, we only downscale 3 ensemble members per lead time and per event. To choose these members out of the 50 members made available by ECMWF, we pick the target ensemble members in order to maximize the expected spread of precipitation according to the following procedure: We assume the expected spread of precipitation to be correlated with the precipitation given by each IFS member in the target domain $d_3$. With this in mind, we sort members across ascending levels of precipitation in the target geographical domain. We depict the ordering of these ensemble members for each event and each lead time in Figure~\ref{figs:ordered-spread}. Observe that, because we order ensemble members per mean precipitation, monotonically increasing curves are displayed. We then pick the ensemble members to maximize the spread by choosing the one with the lowest average precipitation (member 1), the one with the highest (member 3) and the one in the middle (member 2).

\subsection{Generative Downscaling}
\label{sec:generative-downscaling}

For our generative downscaling method, we use generative diffusion models \cite{croitoru2023diffusion}. More specifically, we adapt the diffusion bridge framework from \citeA{bischoff2024unpaired} to subseasonal precipitation downscaling as it is unpaired, i.e., it does not require IFS data for training. This makes the comparison to dynamical downscaling fairer because a single trained ML model can in principle be used to downscale any forecast. Samples are generated from noised images that are sequentially ``denoised" through a stochastic differential equation (SDE), described in Section~\ref{sec:diffusion-models}. To add information from the original low-scale forecasts, we use a posterior conditioning method, described in Section~\ref{sec:posteriori-conditioning}. We depict pre-processing steps applied to the forecasts in Section~\ref{sec:pre-processing-steps}.

\subsubsection{Denoising Diffusion Probabilistic Models}
\label{sec:diffusion-models}

Following \citeA{song2020score}, a diffusion model can be conceived as bridging an unknown distribution density $p(\boldsymbol{x}_0)$ underlying a random variable $\boldsymbol{X}_0$ (representing our target high-resolution data) and a known distribution density $p(\boldsymbol{x}_1)$ underlying a random variable $\boldsymbol{X}_1$ (which is constructed to be a Gaussian). To move in this bridge, we consider an artificially created time variable $t \in [0,1]$. The forward path is

\begin{equation}
\label{eq:forward}
\mathrm{d}\boldsymbol{X}_t = g(t)\,\mathrm{d}\boldsymbol{W}_t,
\end{equation}
where $g$ is the diffusion function, $\boldsymbol{X}_t$ is the distribution at time $t$ (which has density $p(\boldsymbol{x}_t)$), and $\boldsymbol{W}_t$ is the Wiener process (Brownian motion). Given the initial $\boldsymbol{X}_0$, the solution can be written as

\begin{equation}
 \boldsymbol{X}_t=\boldsymbol{X}_0 + \boldsymbol{N}_t,
\end{equation}
where $\boldsymbol{N}_t \sim \mathcal{N}(\boldsymbol{0},\sigma(t)^2)$ and
\begin{equation}
 g(t)^2=\sigma(t)\frac{\mathrm{d}\sigma(t)}{\mathrm{d}t},
\end{equation}
for a smooth function $\sigma(t)$ \cite{sarkka2019applied}. This function is prescribed such that, as $t$ grows, the original image is gradually corrupted until its original information is approximately completely erased at $t=1$.

As the diffusion coefficient depends only on time, \citeA{anderson1982reverse} showed that the reverse process follows

\begin{equation}
\label{eq:backward}
\mathrm{d}\boldsymbol{X}_t = - g(t)^2 \nabla \log p(\boldsymbol{X}_t)\,\mathrm{d}\bar{t} + g(t)\,\mathrm{d}\bar{\boldsymbol{W}}_t,
\end{equation}
where $\mathrm{d}\bar{t}$ is a negative time increment (going backward from $t=1$ to $t=0$), $\mathrm{d}\bar{\boldsymbol{W}}_t$ is the reverse Wiener process, and $\nabla \log p(\boldsymbol{x}_t)$ is the score function. By estimating the score, one can cross the bridge backwards, from $\boldsymbol{X}_1$ at $t=1$ (Gaussian noise) to $\boldsymbol{X}_0$ at $t=0$ (clean data). For complex distributions, such as the ones in the images provided by weather forecasts, we learn the score with a neural network $s_{\boldsymbol{\theta}}(\boldsymbol{x}_t,t) \approx \nabla \log p(\boldsymbol{x}_t)$ with flexible parameters $\boldsymbol{\theta}$. For more details on the diffusion function governing the SDEs, see \ref{app:diffusion-function}; we describe the practical details for sampling images from the backward SDE in \ref{app:sampling-procedure}; the training details to learn the score function can be found in \ref{app:training-details}; the architecture employed for the score $s_{\boldsymbol{\theta}}$ is described in \ref{app:score-architecture}.

\subsubsection{Conditioning a Posteriori}
\label{sec:posteriori-conditioning}

\begin{figure}
 \centering
 \includegraphics[width=\linewidth]{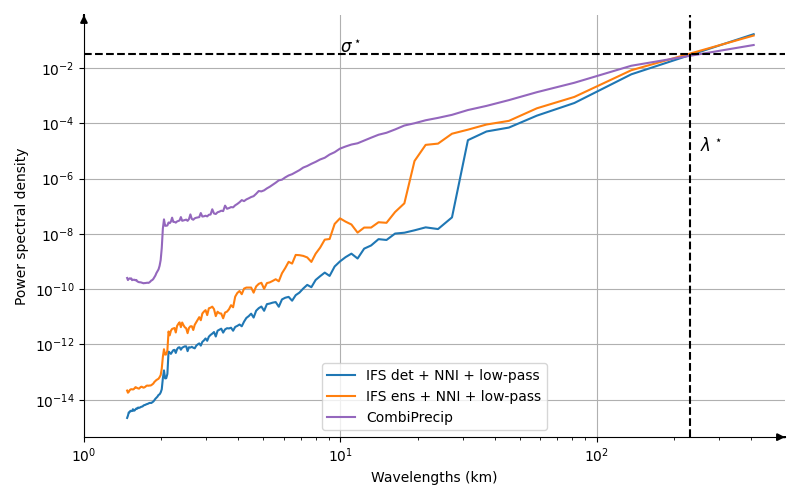}
 \caption{Power spectral density (PSD) of the two pre-processed forecast products from IFS to be used as conditions a posteriori by the diffusion model and the ground truth data provided by CombiPrecip. The pre-processing consists of a nearest neighbors interpolation (NNI) followed by a low-pass filter. The lines show the PSD of the 3-week-ahead forecasts. The 1-week-ahead and 2-week-ahead forecasts look qualitatively similar. The two dashed lines show the optimal standard deviation Gaussian noise $\sigma^\star$ and the associated optimal wavelength $\lambda^\star$ up to which we noise/denoise the forecasts using the diffusion bridge.}
 \label{figs:psd-tuning}
\end{figure}

The key task in downscaling is to incorporate information from low-resolution forecasts while generating high-resolution outputs. We achieve this using the method proposed in \citeA{bischoff2024unpaired}. The conditions are imposed after the model is trained (a posteriori), hence one single trained diffusion model may be used to downscale any forecast. This is done by only integrating the forward pass up to some

\begin{equation}
 \sigma^\star \equiv \sigma(t^\star) = \sqrt{n_x n_y \cdot \text{PSD}_{\boldsymbol{X}_0}\left(\frac{2\pi}{\lambda^\star}\right)},
\end{equation}
where $n_x$ and $n_y$ are the number of pixels in each axis of the image. Above, the noise level $\sigma^\star$ under the associated time $t^\star$ is calculated for some $\lambda^\star$ representing a target scale where low resolution images become indistinguishable to high resolution images from the spectral point of view. In other words, we assume that the PSD of the low-resolution and the high-resolution images match for scales larger than $\lambda^\star$ (the resolvable scales of the low-resolution data), but diverge at smaller scales (the unresolved scales we want to generate). As a consequence, one can use the low-resolution distribution in the forward pass from $t=0$ to $t=t^\star$ and go back to $t=0$ using the score learned from high resolution distribution, which should correct the low-resolution images with the small scale features learned from the high-resolution ones. In practice, $\lambda^\star$ is chosen to be a point after which scales match approximately. We provide the spectral analysis used for the tuning of the $\lambda^\star$ parameter in Figure~\ref{figs:psd-tuning}. Shown is the PSD of the two pre-processed IFS forecast products (which are not used during training), as well as the PSD of the high-resolution dataset provided by CombiPrecip (the one used to train the model). The pre-processing consists of a nearest neighbor (NNI) interpolation and a low-pass filter, which are detailed in the next section. We take $\lambda^\star$ around $200~\unit{km}$, although another possible choice would be to take it around $30~\unit{km}$, but we have chosen the former because it led to better resolved small scales. Indeed, with more noise, the diffusion model can have more denoising steps and produce generally better results, although becoming less attached to the original low-resolution information. Moreover, we use the same $\lambda^\star$ for all lead times and IFS products (deterministic and ensemble) as they all look very similar. A more principled criterion for selecting such a parameter is left for future work.

By doing so, we are conditioning the generation described in the previous section a posteriori, i.e., after training. As a consequence, the trained model is unpaired from the original forecast to be downscaled. Indeed, only the high-resolution data (CombiPrecip) is used to train the model and the low-resolution data (coming from the IFS forecasts) is used to condition the generation of the already-trained model after the tuning parameter $\lambda^\star$ is established. This makes the diffusion model more general and goes in hand with a fairer comparison with the general-purpose WRF model.

\subsubsection{Pre-Processing Steps}
\label{sec:pre-processing-steps}
As our generative models and evaluation metrics need to operate at the same resolution as the ground truth data, we apply pre-processing steps to the low-resolution dataset. First, we interpolate it to match the high-resolution dataset grid. This is done using a nearest-neighbor (NNI) method, which replicates the original image with more grid points. We also use this interpolation to evaluate the original skill of the low-resolution datasets, including the non-downscaled 0-day forecast. Before feeding the interpolation to the diffusion model, we apply a low-pass filter to remove artificially created high-frequency content, using the Nyquist wavenumber of the original low-resolution dataset as a cutoff.

Let $I(x,y)$ denote an image, where $x \in \{0,1,...,n_x-1\}$ and $y \in \{0,1,...,n_y-1\}$ with $n_x$ and $n_y$ representing the high-resolution image dimensions. To maintain dimensional consistency with physical parameters, we express the discrete Fourier transform (DFT) using integer frequency indices $u$ and $v$ like

\begin{equation}
 \tilde{I}(u,v) = \sum_{x=0}^{n_x-1}\sum_{y=0}^{n_y-1}I(x,y) \mathrm{e}^{-2\pi i\left(\frac{ux}{n_x} + \frac{vy}{n_y}\right)},
\end{equation}
where $u \in \left\{-\frac{n_x}{2}, \dots, \frac{n_x}{2}-1\right\}$ and $v \in \left\{-\frac{n_y}{2}, \dots, \frac{n_y}{2}-1\right\}$. These integer indices correspond directly to the physical angular wavenumbers via $k_x = \frac{2\pi u}{n_x \Delta x}$ and $k_y = \frac{2\pi v}{n_y \Delta y}$, where $\Delta x$ and $\Delta y$ represent the physical grid spacing of the high-resolution image. 

\begin{figure}
 \centering
 \includegraphics[width=\linewidth]{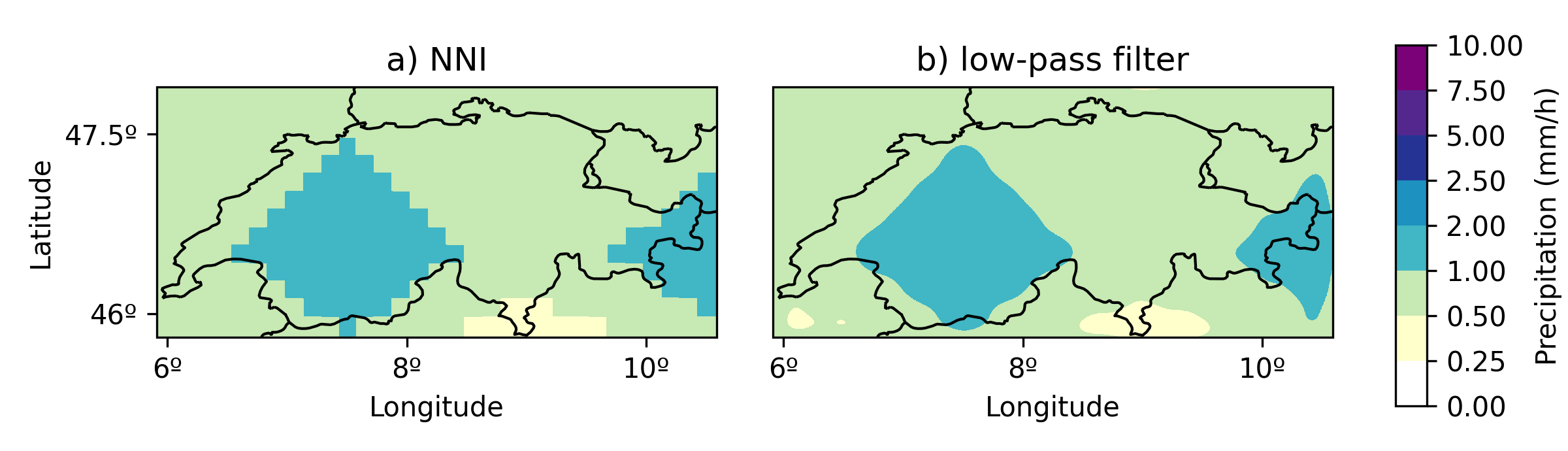}
 \caption{Pre-processing steps applied to the raw IFS forecast before feeding them as conditions to the diffusion model. (a) The nearest neighbor  interpolation (NNI), which maintains the same information present in the raw forecast, but increases the number of points in the image to the high-resolution reference. (b) The radial low-pass filter used to take out spurious edges remaining from the interpolation.}
 \label{figs:pre-processing}
\end{figure}

We compute this DFT after interpolation. We then reconstruct the image with the inverse DFT, but only keep frequencies smaller than the Nyquist limit of the original low-resolution dataset. In terms of our discrete frequency indices, the low-resolution cutoffs  elegantly simplify to $u^c = \frac{n_x^\text{low-res.}}{2}$ and $v^c = \frac{n_y^\text{low-res.}}{2}$, where $n_x^\text{low-res.}$ and $n_y^\text{low-res.}$ represent the original low-resolution image dimensions. The resulting filtered image $I'$ can be written as

\begin{equation}
 I'(x,y) 
 = \frac{1}{n_x n_y} 
 \sum_{u=-\frac{n_x}{2}}^{\frac{n_x}{2}-1} 
 \sum_{v=-\frac{n_y}{2}}^{\frac{n_y}{2}-1} 
 \tilde{I}(u,v) 
 \mathrm{e}^{2\pi i \left( \frac{ux}{n_x} + \frac{vy}{n_y} \right)} 
 \mathbbm{1}_{\left\{|u| \leq u^c \right\}} 
 \mathbbm{1}_{\left\{|v| \leq v^c \right\}},
\end{equation}
where $\mathbbm{1}$ represents the indicator function (equal to 1 when the condition is true, 0 otherwise). This produces a more realistic interpolation by removing spurious high-frequency content introduced by the nearest-neighbor interpolation process. This filtered result serves as the starting point for the diffusion-based downscaling model. The pre-processing step-by-step can be seen in Figure~\ref{figs:pre-processing}.

\section{Results}
\label{sec:results}
Here we present the results of our experiments. Throughout the figures and tables displayed in this section, we compare the original forecasts provided by IFS from ECMWF, the diffusion model ensembles serving as the generative downscaling model, and WRF serving as the dynamical downscaling model against the ground truth provided by CombiPrecip. In our experiments, IFS~det~+~NNI and IFS~ens~+~NNI refer to the original deterministic and 50-member ensemble subseasonal forecasts provided by IFS, respectively, plus an NNI interpolation. The NNI interpolation is carried out for ease of comparison, as it brings the low-resolution forecasts to the common downscaled high resolution, even though the forecasts visually look like they have the same low resolution as before. DDPM~det and DDPM~ens refer to the ensembles generated using the diffusion model conditioned on the deterministic and the ensemble forecast, respectively, as the conditional information. For DDPM~det, one image is used as a condition to generate 50 samples for each time frame---the original deterministic image from ECMWF. For DDPM~ens, each of the 50 members in the original IFS ens forecast is used as a condition for a single sample generation using the diffusion model---this is done so that the two diffusion sample ensembles have the same number of members. As a consequence, DDPM~det has an ensemble that underlies the variability in samples of the diffusion model, which is, with some training approximations, the variability of the training set provided by CombiPrecip, as discussed in Section~\ref{sec:data}. The DDPM~ens inherits not only from the variability of CombiPrecip, but also from the ensemble generated by IFS. We remind the reader that both DDPM~det and DDPM~ens are diffusion model samples using one single trained neural network, but under two different sampling procedures. Finally, as detailed in Section~\ref{sec:dynamical-downscaling}, we generate 3 ensemble members using the IFS ensemble forecast (with a wider geographical domain and more variables) for WRF. Even though this is a small ensemble, it is a direct consequence of the computational cost of the model. The summary of each model is given in Table~\ref{tab:model-summary}.

\begin{table}
 \centering
 \begin{tabular}{lllc}
 \hline
 \textbf{Model} & \textbf{Type} & \textbf{Conditioning} & \textbf{Ensemble Size} \\
 \hline
 CombiPrecip & Ground truth & -- & 1 \\
 IFS~det~+~NNI & Original forecast & -- & 1 \\
 IFS~ens~+~NNI & Original forecast & -- & 50 \\
 DDPM~det & Generative downscaling & IFS det & 50 \\
 DDPM~ens & Generative downscaling & IFS ens & 50 \\
 WRF & Dynamical downscaling & IFS ens* & 3 \\
 \hline
 \end{tabular}
 \caption{Summary of the models evaluated in the experiments, including their type, conditioning data, and ensemble size. The ground truth (CombiPrecip) and the forecast products provided by ECMWF (IFS~det~+~NNI and IFS~ens~+~NNI) are described in Section~\ref{sec:data}. The diffusion model (DDPM~det and DDPM~ens; generative downscaling) is described in Section~\ref{sec:generative-downscaling}. WRF (dynamical downscaling) is described in Section~\ref{sec:dynamical-downscaling}. IFS ens* refers to the fact that the boundary conditions used by WRF lie in a wider geographical domain with many more variables than the single and restricted precipitation field provided as an input to the diffusion model.}
 \label{tab:model-summary}
\end{table}

\begin{figure}
 \centering
 \includegraphics[width=\linewidth]{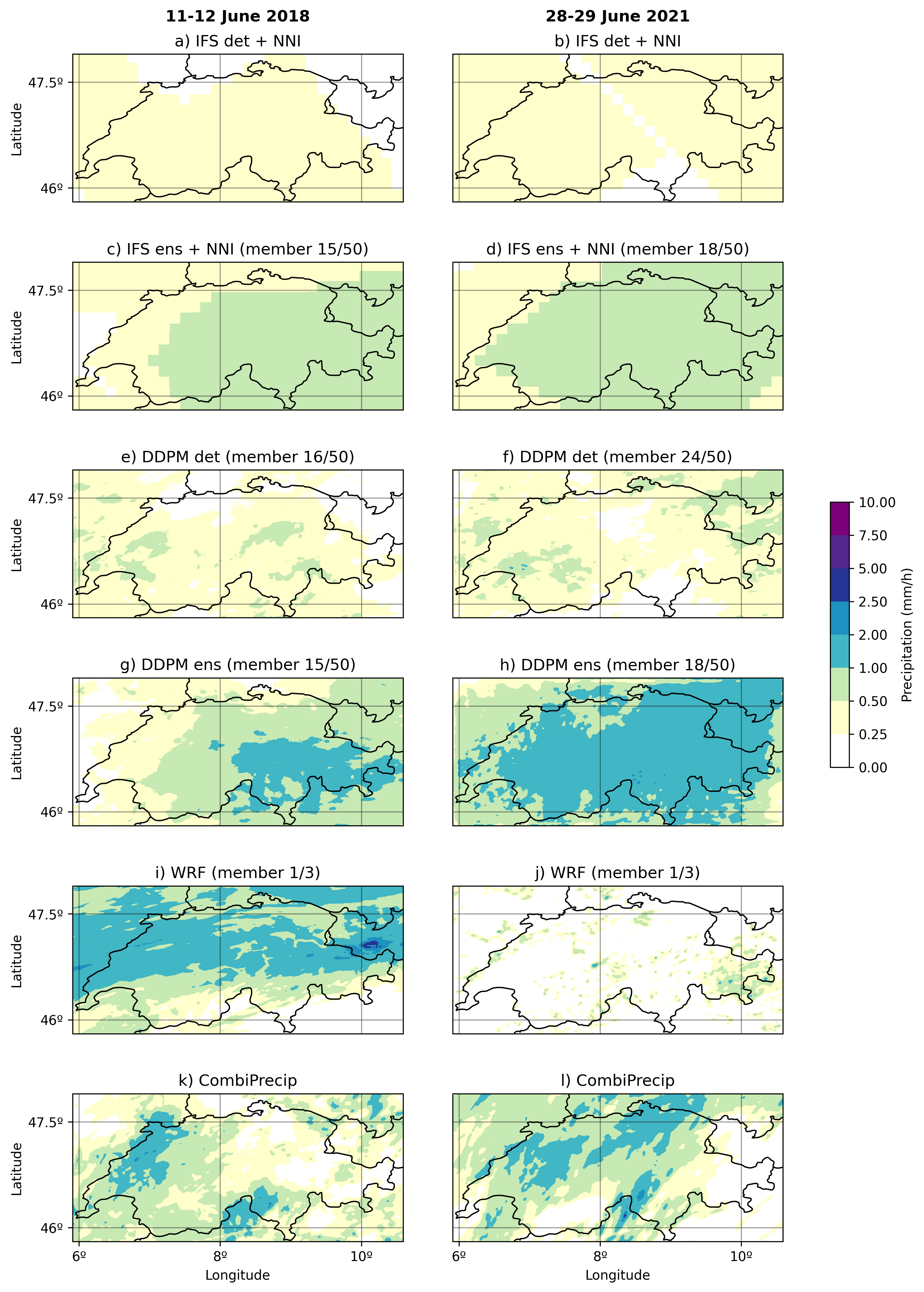}
 \caption{(a-j) Mean aggregated precipitation for all five benchmarked models compared against the (k-l) CombiPrecip ground truth at a 3-week lead time for the 2018 (left column) and 2021 (right column) extreme precipitation events. Rows from top to bottom display (a-b) IFS det + NNI, (c-d) IFS ens + NNI, (e-f) DDPM det, (g-h) DDPM ens, (i-j) WRF. For ensemble models, the displayed member corresponds to the one with the highest total accumulated precipitation over the event window.}
 \label{figs:maps-3-week}
\end{figure}

We provide a visual snapshot of each model in comparison with the ground truth provided by CombiPrecip in Figure~\ref{figs:maps-3-week} for the event on 11-12 June 2018 (on the left) and on 28-29 June 2021 (on the right) averaged for the 48~h of the event with the 3-week lead time. In the figure, the chosen ensemble members display the greatest overall mean precipitation among all ensemble members for each model, lead time and event. We additionally provide the snapshots for the 1-week and 2-week lead times in Figure~\ref{figs:maps-1-week}-\ref{figs:maps-2-week}, respectively.

The two subseasonal forecasts (Figure~\ref{figs:maps-3-week}a-d) visually display lower resolution, despite the NNI interpolation, while the samples from the diffusion models (Figure~\ref{figs:maps-3-week}e-h), the WRF simulation (Figure~\ref{figs:maps-3-week}i-j) and the ground truth (Figure~\ref{figs:maps-3-week}k-l) display more spatial variability on the scales close to the $1$~\unit{km}-resolution. As mentioned previously, the diffusion models are conditioned to use the low-resolution data provided by the original forecast constrained to the Swiss region in the map, hence inheriting information from  large-scale circulation features, but can bias-correct and augment the resolution of the lower scales. Indeed, one can infer the resemblance between IFS det~and DDPM~det and between IFS ens~and DDPM~ens. Moreover, even though WRF is driven by the same subseasonal forecast, it uses multiple variables, including variables at different pressure levels, and has boundary conditions coming from a wider spatial domain in comparison to the region displayed in the figures.

\begin{table}
\centering
\begin{tabular}{l cccc cccc}
\hline
& \multicolumn{4}{c}{2018 event} & \multicolumn{4}{c}{2021 event} \\
\cmidrule(lr){2-5} \cmidrule(lr){6-9}
& 0-day & 1-week & 2-weeks & 3-weeks & 0-day & 1-week & 2-weeks & 3-weeks \\
\midrule
IFS~det~+~NNI & -0.03 & 0.00 & 0.02 & -0.04 & 0.12 & 0.00 & -0.12 & -0.12 \\
IFS~ens~+~NNI & 0.17 & 0.21 & \textbf{0.20} & 0.18 & 0.28 & \textbf{0.15} & \textbf{0.13} & \textbf{0.17} \\
DDPM~det & - & 0.11 & 0.05 & 0.04 & - & 0.00 & -0.16 & -0.03 \\
DDPM~ens & - & 0.23 & 0.19 & 0.18 & - & \textbf{0.15} & \textbf{0.13} & \textbf{0.17} \\
WRF & - & \textbf{0.27} & 0.14 & \textbf{0.32} & - & 0.01 & 0.01 & 0.04 \\
\hline
\end{tabular}
\caption{Continuous Ranked Probability Score (CRPS) skill score with respect to the climatology for 0-day, 1, 2, and 3-week lead times for the two events. Higher values are better and the best model(s) for each column are highlighted in bold.}
\label{tab:metrics-crpsss}
\end{table}

The performance of the forecast models is evaluated for lead times of 1, 2, and 3 weeks. The results for the CRPS skill score are shown in Table~\ref{tab:metrics-crpsss}, W1 skill score in Table~\ref{tab:metrics-w1ss}, avPSDd skill score in Table~\ref{tab:metrics-avpsddss}, and the SSR in Table~\ref{tab:metrics-ssr}. The forecasting skill for the 0-day forecast is also provided for comparison. In Table~\ref{tab:metrics-crpsss}, Table~\ref{tab:metrics-w1ss} and Table~\ref{tab:metrics-avpsddss}, higher values represent better performance, and the highest-scoring model for each lead time is highlighted in bold. In Table~\ref{tab:metrics-ssr},  values closer to 1 indicate better performance.

\begin{table}
\centering
\begin{tabular}{l cccc cccc}
\hline
& \multicolumn{4}{c}{2018 event} & \multicolumn{4}{c}{2021 event} \\
\cmidrule(lr){2-5} \cmidrule(lr){6-9}
& 0-day & 1-week & 2-weeks & 3-weeks & 0-day & 1-week & 2-weeks & 3-weeks \\
\midrule
IFS~det~+~NNI & 0.42 & 0.34 & 0.08 & 0.39 & 0.55 & -0.09 & 0.61 & 0.43 \\
IFS~ens~+~NNI & 0.45 & 0.49 & 0.36 & 0.31 & 0.55 & 0.22 & 0.19 & 0.24 \\
DDPM~det & - & 0.44 & 0.03 & 0.50 & - & -0.08 & \textbf{0.85} & \textbf{0.50} \\
DDPM~ens & - & 0.66 & \textbf{0.43} & 0.36 & - & \textbf{0.27} & 0.23 & 0.31 \\
WRF & - & \textbf{0.88} & 0.18 & \textbf{0.79} & - & -0.05 & 0.05 & -0.01 \\
\hline
\end{tabular}
\caption{Wasserstein distance skill score with respect to the climatology for 0-day, 1, 2, and 3-week lead times for the two events. Higher values are better and the best model for each column is highlighted in bold.}
\label{tab:metrics-w1ss}
\end{table}

Table~\ref{tab:metrics-w1ss} presents the Wasserstein distance skill scores. The best scores vary by event and lead time, and are divided between WRF and the diffusion models. For the 2018 event, WRF achieves the highest scores at the 1-week (0.88) and 3-week (0.79) lead times, while DDPM~ens~leads at the 2-week mark (0.43). For the 2021 event, the diffusion models capture the highest scores. Across both events and all evaluated lead times, both diffusion models (DDPM~det and DDPM~ens) generally give higher skill scores than their corresponding IFS forecasts (IFS~det~+~NNI and IFS~ens~+~NNI).

\begin{table}
\centering
\begin{tabular}{l cccc cccc}
\hline
& \multicolumn{4}{c}{2018 event} & \multicolumn{4}{c}{2021 event} \\
\cmidrule(lr){2-5} \cmidrule(lr){6-9}
& 0-day & 1-week & 2-weeks & 3-weeks & 0-day & 1-week & 2-weeks & 3-weeks \\
\midrule
IFS~det~+~NNI & 0.02 & 0.02 & 0.18 & -0.05 & 0.02 & \textbf{0.15} & -0.02 & -0.50 \\
IFS~ens~+~NNI & 0.03 & 0.10 & -0.05 & -0.12 & \textbf{0.10} & -0.16 & -0.11 & -0.18 \\
DDPM~det & - & 0.27 & \textbf{0.41} & 0.22 & - & -0.71 & 0.11 & -0.17 \\
DDPM~ens & - & 0.29 & 0.15 & 0.09 & - & 0.00 & 0.06 & \textbf{0.01} \\
WRF & - & \textbf{0.49} & 0.18 & \textbf{0.23} & - & 0.01 & \textbf{0.21} & -0.34 \\
\hline
\end{tabular}
\caption{Member-averaged power spectral density distance skill score (avPSDd) with respect to the climatology for 0-day, 1-, 2-, and 3-week lead times for the two events. To avoid introducing variance artifacts from flat fields, the $0.1$~\unit{mm/h} trimming threshold applied to other metrics is omitted for the PSD evaluations. Higher values are better and the best model of each column is highlighted in bold.}
\label{tab:metrics-avpsddss}
\end{table}

Table~\ref{tab:metrics-avpsddss} presents the avPSDd skill scores. Note that to avoid variance artifacts associated with thresholding low-intensity precipitation, the $0.1$~\unit{mm/h} trimming was omitted explicitly for this metric. Here, the diffusion models (DDPM~det and DDPM~ens) and WRF generally maintain an advantage over the baseline provided by ECMWF. For the 2018 event, WRF demonstrates the highest skill at the 1-week (0.49) and 3-week (0.23) lead times, while the deterministic diffusion model (DDPM~det) achieves the best score at the 2-week mark (0.41). Conversely, the 2021 event proves more challenging to reconstruct, with models displaying negative skill scores (hence worse than the climatology) more often. IFS~det~+~NNI records the highest skill at the 1-week lead time (0.15), WRF leads at the 2-week mark (0.21), and DDPM~ens~achieves the best score at 3 weeks (0.01).

\begin{figure}
 \centering
 \includegraphics[width=\linewidth]{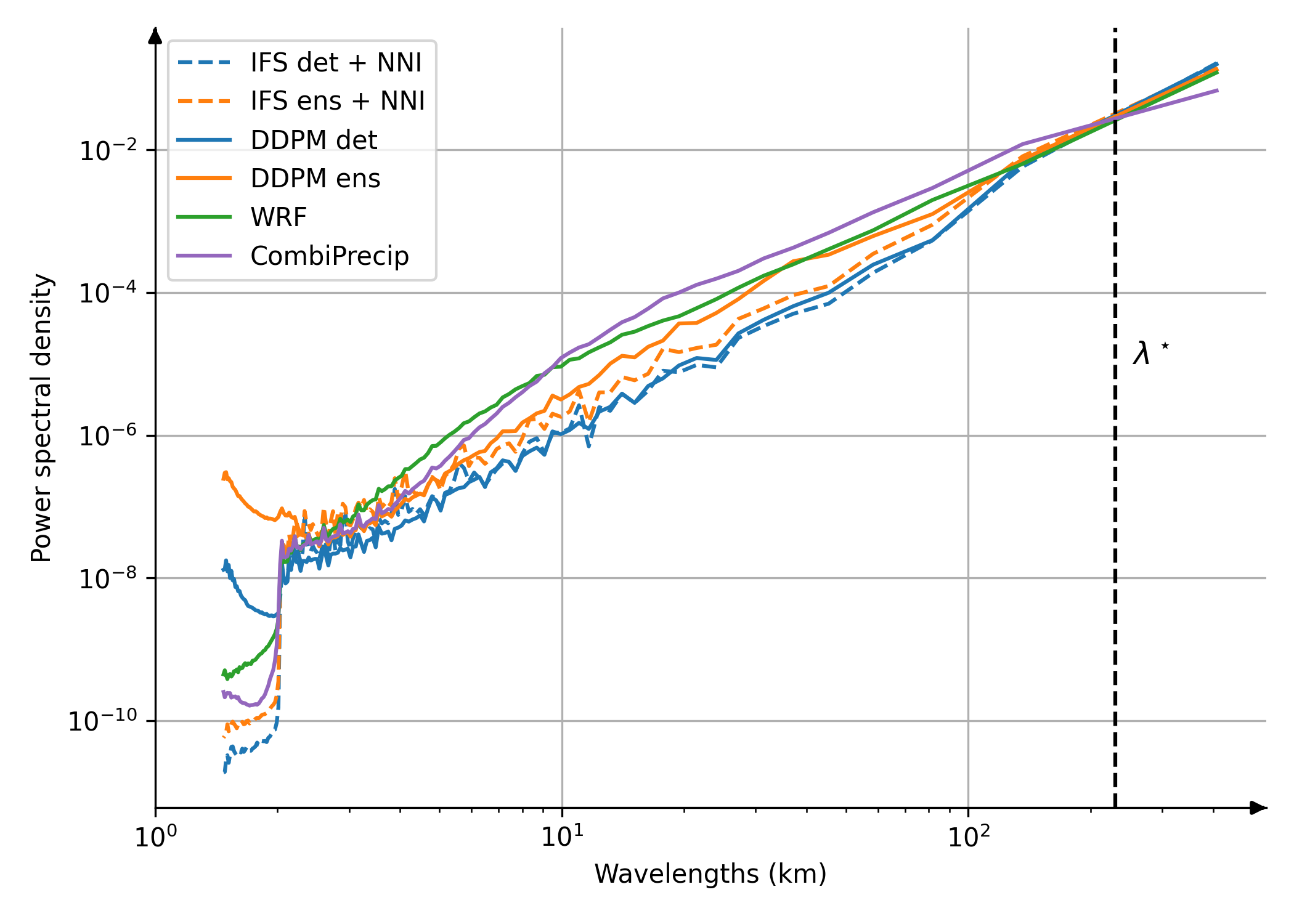}
 \caption{Power spectral density of the 5 benchmarked models for the 3-week lead time as well as the ground truth provided by CombiPrecip. The two events are concatenated over the time axis to produce the spectrum.}
 \label{figs:psd-3-week}
\end{figure}

Figure~\ref{figs:psd-3-week} shows the PSD as a function of  wavelength in kilometers for the forecasts with a lead time of 3 weeks, where the two events are concatenated over the time axis following the same aggregation done for simplification in Section~\ref{sec:posteriori-conditioning}. We additionally provide the PSD corresponding to lead times of 1  and 2 weeks in Figure~\ref{figs:psd-1-week} and Figure~\ref{figs:psd-2-week}, respectively. All models diverge at scales lower than 2~km, close to the resolution of CombiPrecip. Moreover, the diffusion model is very close to the original IFS condition for scales higher than $\lambda^\star$, but it adds information for lower scales, up to the training quality of denoising images.

\begin{table}
\centering
\begin{tabular}{l cccc cccc}
\hline
& \multicolumn{4}{c}{2018 event} & \multicolumn{4}{c}{2021 event} \\
\cmidrule(lr){2-5} \cmidrule(lr){6-9}
& 1-day & 1-week & 2-weeks & 3-weeks & 1-day & 1-week & 2-weeks & 3-weeks \\
\midrule
IFS~det~+~NNI & - & - & - & - & - & - & - & - \\
IFS~ens~+~NNI & 0.59 & 0.44 & 0.45 & 0.40 & 0.57 & 0.37 & 0.31 & 0.39 \\
DDPM~det & - & 0.22 & 0.13 & 0.22 & - & 0.03 & 0.20 & 0.20 \\
DDPM~ens & - & 0.66 & \textbf{0.65} & 0.60 & - & \textbf{0.56} & \textbf{0.44} & \textbf{0.62} \\
WRF & - & \textbf{0.98} & 0.59 & \textbf{1.04} & - & 0.14 & 0.34 & 0.29 \\
\hline
\end{tabular}
\caption{Spread-skill ratio (SSR) for 0-day, 1, 2, and 3-week lead times for two events. Values closer to 1 indicate better probabilistic calibration, and the best model of each column is highlighted in bold. Values for IFS~det~+~NNI are omitted (-) because it is a deterministic forecast.}
\label{tab:metrics-ssr}
\end{table}

The spread-skill ratio (SSR) is presented in Table~\ref{tab:metrics-ssr}. Values below 1 indicate under-dispersion, which is prevalent across most models and lead times. The deterministic diffusion model (DDPM~det) is the most under-dispersed overall. However, the diffusion ensemble (DDPM~ens) consistently improves the spread of the original IFS ensemble (IFS~ens~+~NNI), as it adds the variability of the diffusion model to the variability of the perturbed forecast. As the ECMWF perturbations are under-dispersed, DDPM~ens~yields better SSR values in comparison to IFS~ens~+~NNI. Moreover, DDPM~ens~yields the best spread for all lead times for the 2021 event. WRF demonstrates the best overall calibration for the 2018 event, with SSR values of 0.98 and 1.04 for 1 week and 3 weeks of lead time, respectively. Furthermore, IFS ens~exhibits the  highest skill for the 0-day lead time and plateaus for the 1-week, 2-week and 3-week lead times for both events.

\begin{figure}
 \centering
 \includegraphics[width=\linewidth]{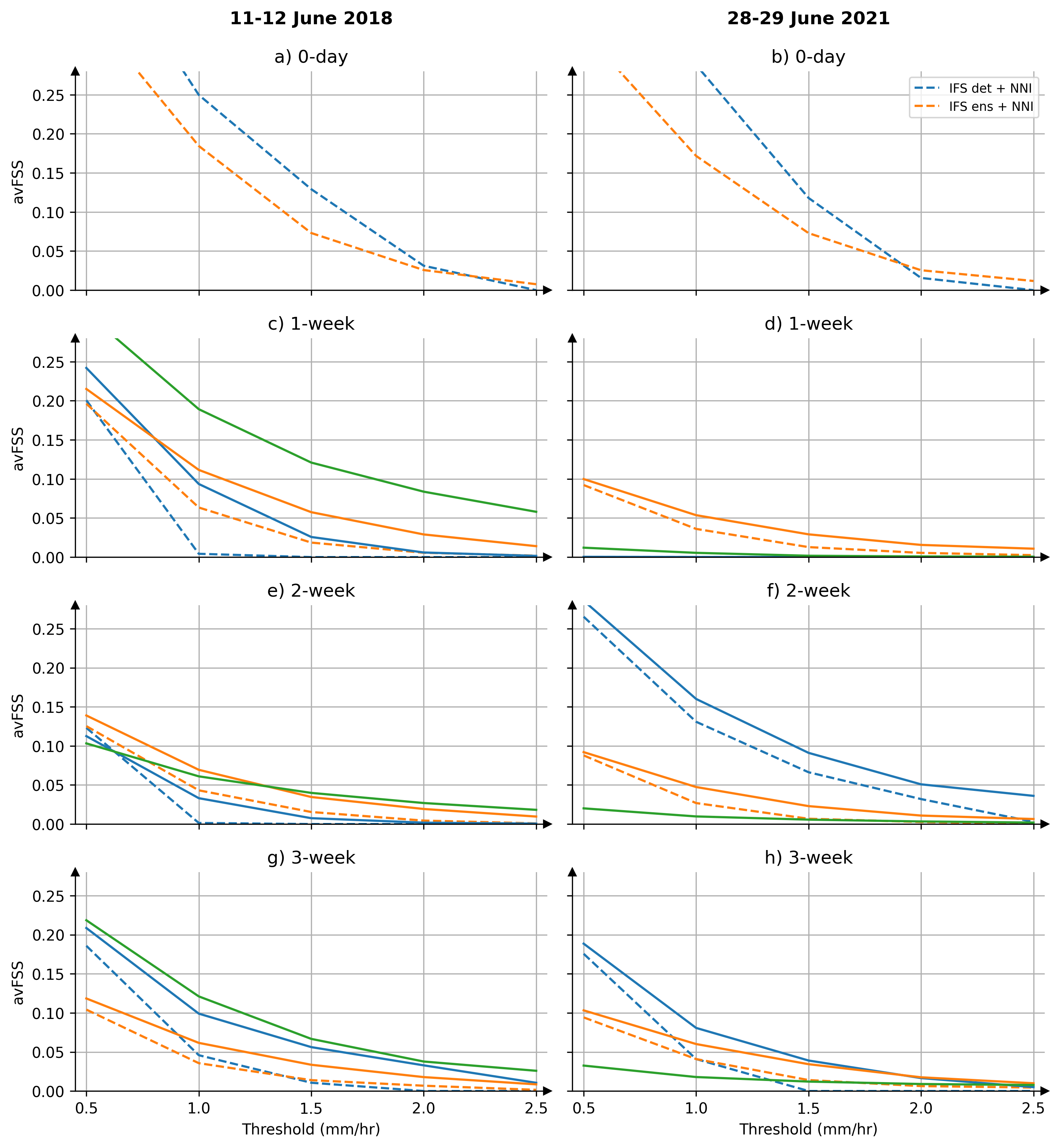}
 \caption{Fraction skill score avFSS for all models and for all lead times as a function of the precipitation threshold using CombiPrecip as ground truth for the event on 11-12 June 2018 (on the left) and on 28-29 June 2021 (on the right) for the lead times of (a-b) 0 day, (c-d) 1 week, (e-f) 2 weeks, and (g-h) 3 weeks.}
 \label{figs:fss}
\end{figure}

The member-averaged FSS is computed using 15 pixels (hence $15~\unit{km} \times 15~\unit{km}$ since the resolution of the downscaled forecasts is $1~\unit{km} \times 1~\unit{km}$) as the moving window. This is shown for all models in Figure~\ref{figs:fss} for the event on 11-12 June 2018 (on the left) and on 28-29 June 2021 (on the right). The figure clearly demonstrates the skill added by the diffusion model in relation to the original forecast source, as it consistently improves the avFSS for all thresholds displayed in the figure. WRF also exhibits a better avFSS for higher values of the precipitation threshold in comparison to the other models, showing the value that WRF adds for the extreme bits of the forecasts. We do not include higher threshold values in the plot because the avFSS is close to zero for all the models. In our experiments, changing the moving window size (to 30~\unit{km}, 45~\unit{km}, 60~\unit{km}) does not qualitatively impact these results.

\section{Discussion}
\label{sec:discussion}

We now contrast the results between the multiple models, events and lead times presented in the previous section. To ensure a fair comparison with dynamical downscaling, we train the diffusion model (generative downscaling) in an unpaired fashion. This procedure is fundamentally different from ML models trained to downscale a specific global weather prediction model, which would mostly lead to better performance in our case of downscaling precipitation over Switzerland \cite{assouline2026lead}. Unlike such tailored models, our unpaired model can be used to downscale different products: from reanalysis to subseasonal forecasts, and in theory, even climate projections. By doing so, we not only make the model versatile, but also avoid the common pitfall of ML models being trained to outperform baselines for a narrow range of events or boundary conditions. In fact, even though we use two extreme events to benchmark the models, we have not tailored our ML model to extreme precipitation, which could be done by, for instance, oversampling higher values of precipitation in the dataset. Ultimately, this makes our comparison fairer and potentially closer to a real use-case scenario.

From the standpoint of usability, the primary trade-off between WRF and the diffusion model is the time of calibration and simulation. On the one hand, WRF is a general-purpose physics-based model: once the boundary conditions for the inner domains are set, WRF can be applied to any region of the world without data-driven recalibration. However, in our case, a single ensemble member simulation required approximately $24$~\unit{h} to run on a high-performance cluster using multiple CPU threads. On the other hand, the diffusion model must be trained on data for the specific region and time resolution of interest. This one-time training process took around $10$~\unit{h} on a single NVIDIA A100 GPU. Once trained, generating a 50-member ensemble for each of the three lead times took $10$~\unit{min} when driven by a deterministic forecast, and $20$~\unit{min} when using an ensemble forecast as input---making it thousands of times faster than the physics-based counterpart once trained. This difference exists because it is easier to parallelize samples for a single initial condition in the sampling SDE, even though it should also be possible to do the same for multiple initial conditions. Note that this comparison uses wall-clock time rather than a normalized hardware-cost metric (e.g., CPU- vs. GPU-hours), so the actual resource trade-off may differ.

For the multicellular event on 11-12 June 2018, WRF (used for dynamical downscaling) generally achieves the highest skill scores at the 1-week and 3-week lead times, while, at the 2-week lead time, it is not the top performer but still outperforms climatology across all skill metrics. In fact, for the 2-week lead time, WRF is also not far behind the best model for the CRPS (0.14 vs. 0.20 given by IFS~ens~+~NNI) and for the SSR (0.59 vs. 0.65 given by DDPM~ens). This demonstrates WRF's skill in resolving weather dynamics that are more non-stationary, thanks to its fine inner time-step of $4$~\unit{s}, which better captures the fast evolution of multicellular convection than the driving forecast's coarser temporal resolution allows. Moreover, WRF achieves overall very good dispersion skill for this event, with an SSR of 0.98 (for the 1-week lead time) and 1.04 (for the 3-week lead time), being close to the perfect SSR value of 1. The generative downscaling approach falls short here, mainly because it uses static conditions of the fixed target geographical domain. Despite this, the diffusion models approximately match the CRPS of the original corresponding forecasts from IFS, and consistently improves the probability and spectral distances (W1 and avPSDd) across all lead times. This result demonstrates that, even though the diffusion model is trained in a restricted domain in space and time, it manages to correct spatial structure errors of the raw IFS forecast without losing forecast skill. The improvement with respect to the baseline is clear in Figure \ref{figs:fss}, where the diffusion models consistently score a higher avFSS than their respective conditional fields coming from IFS for all thresholds considered.

For the supercell event on 28-29 June 2021, the subseasonal ensemble forecast (IFS ens) and the associated diffusion model (DDPM~ens) exhibit the highest skill. This is likely because the more coherent, systematic propagation of a passing supercell is already well captured by the driving forecast's coarser temporal resolution, so the diffusion model, which inherits this same time step, adds most of its value by correcting the spatial distribution of precipitation rather than by requiring WRF's finer dynamical time-stepping. Models generally score worse for this event as compared to the 2018 event, except for the 0-day forecasts. The best CRPS for this event is given by the IFS~ens~+~NNI and the corresponding DDPM~ens. For the W1 distance, the diffusion models score the highest, showcasing their skill in probability distributions. In particular, DDPM~det displays good scores in comparison to other models for extended lead times (2-weeks and 3-weeks). For this event, the avPSD distance is split among different models, but more importantly, it is also almost always negative, showing that it is generally difficult to reproduce the small-scale spatial features of a supercell at subseasonal lead times. Regarding the quality of the ensemble dispersion, the models are also generally less dispersed than their counterparts for the event in 2018. Under this metric, DDPM~ens consistently scores as the best model, largely due to the fact that it adds the dispersion of the IFS ensemble to the one inherited by the CombiPrecip training data in an event with overall under-dispersive models. As for the 2018 event, avFSS is consistently higher for the diffusion models in comparison with the raw IFS fields.

In this work, we do not see a clear degradation in performance with increasing lead times. Instead, it is rather the nature of the event and model type that shapes the prediction skill of precipitation for extended lead times. This is largely because extreme precipitation is difficult to predict at lead times shorter than 1-week, so the models already operate close to the climatological limit, e.g., with a relatively small CRPS skill score. Even the 0-day lead time forecasts already suffer from being hours from the thunderstorm events in this study, since the rainfall peak do not occur immediately after the initialization of the evaluation window, which is why the subseasonal downscaling task under comparison here is a genuine stress-test.

Overall, while the diffusion ensemble matches the CRPS of the raw subseasonal ensemble, indicating it preserves the general forecast probabilistic skill at the 1-\unit{km} scale, the model consistently outperforms the raw IFS forecasts in the W1 and avPSDd metrics. This demonstrates that the generative downscaling approach is successfully injecting realistic spatial structure into the precipitation fields. This is also true for the avFSS skill score, which additionally suggests that the downscaling products have a better larger-scale skill than the raw forecasts, as a moving-window of 15-\unit{km} is used to evaluate the models. Ultimately, while only two events are considered, the results align with the distinct physical nature of the two events: WRF excels when an event's non-stationary dynamics evolve faster than the driving forecast's temporal resolution can resolve, requiring the finer dynamical time-stepping of a model like WRF (2018 multicell event), whereas generative downscaling robustly improves forecasts when the driving forecast's own time step already captures the event's dynamics, leaving spatial structure as the main remaining limitation (2021 supercell event).

\section{Conclusion}
\label{sec:conclusion}

In this work, we stress-test dynamical and generative downscaling methods in the challenging task of downscaling subseasonal forecasts of summer convective precipitation events. We compare their ensemble-generation ability, testing both models against two extreme precipitation events in Switzerland. Evaluation against observations using proper scores and spatial, distributional, and spectral diagnostics reveals distinct strengths and trade-offs for both approaches.

In summary, using downscaling models generally improves raw subseasonal forecasts, which motivates their adoption in real-world forecasting pipelines. Diffusion models are better suited for this task due to their lower computational cost and more robust spatial and probability distribution corrections, whereas WRF remains preferable for faster, more non-stationary convection requiring finer time-stepping.

Looking forward, the integration of variables that exhibit intricate interactions with precipitation, such as soil moisture and topography, can further enhance the predictive ability of the generative downscaling model. Soil moisture has shown potential to enhance the predictability of longer events because it can trigger atmospheric instabilities through convection and has a longer memory compared to most atmospheric variables \cite{shin2024deep}. Topography is widely used as a context variable in downscaling models due to its critical role in shaping convective events and its availability at high resolution \cite{tesfa2020exploring, sideris2020nowprecip}. Even though the diffusion model already implicitly accounts for static patterns, which is why we decided not to include static surface predictors here, one can think of ways to use different parts of the geographical domain or augment the dataset in order to insert this information. Moreover, one can also think of using wider geographical domains to train the diffusion model and expect that it could learn how to integrate some synoptic-scale dynamics.

\section*{Open Research Section}
Subseasonal prediction data from ECMWF was not open source at the date of last access (14 November 2025). The WRF model is freely available for download at NCAR's website (\url{https://www.mmm.ucar.edu/models/wrf}, last accessed 13 September 2024). Our code is available at \url{https://doi.org/10.5281/zenodo.22687888}.


\acknowledgments
ML and TB acknowledge support from the Swiss National Science Foundation (SNSF) under Grant No. 10001754 (``RobustSR'' project). ML acknowledges Sorbonne University and Ecole Polytechnique for making this research possible through an UNIL research internship. ML acknowledges the authors of \url{https://github.com/google-research/swirl-dynamics} for their code on diffusion models. We thank MeteoSwiss for providing the data used to train and evaluate the models and ECMWF for providing easily accessible forecast data for the low-resolution baselines. We thank Nadav Peleg for helpful advice regarding WRF downscaling. 
\bibliography{bibliography}

\clearpage

\appendix

\section{WRF inputs}

The table below describes the inputs used by the WRF model, which are made available by the subseasonal prediction product from IFS.

\begin{table}
\centering 
\begin{tabular}{lll}
\textbf{Variable Name} & \textbf{Description} & \textbf{Units} \\
\hline
z & Geopotential & \unit{m} \\
u & Zonal wind & \unit{m/s} \\
v & Meridional wind & \unit{m/s} \\
t & Temperature & \unit{K} \\
q & Specific humidity & \unit{kg/kg} \\
r & Relative humidity & \unit{-} \\
ci & Cloud ice water content & \unit{kg/m^2} \\
rsn & Snow density & \unit{kg/m^3} \\
sstk & Sea surface temperature & \unit{K} \\
swvl1 & Soil moisture layer 1 & \unit{m^3/m^3} \\
swvl2 & Soil moisture layer 2 & \unit{m^3/m^3} \\
swvl3 & Soil moisture layer 3 & \unit{m^3/m^3} \\
swvl4 & Soil moisture layer 4 & \unit{m^3/m^3} \\
sp & Surface pressure & \unit{Pa} \\
tcw & Total cloud water content & \unit{kg/m^2} \\
tcwv & Total column water vapor & \unit{kg/m^2} \\ 
stl1 & Soil temperature layer 1 & \unit{K} \\
stl2 & Soil temperature layer 2 & \unit{K} \\
stl3 & Soil temperature layer 3 & \unit{K} \\
stl4 & Soil temperature layer 4 & \unit{K} \\
sd & Snow depth & \unit{m} \\
msl & Mean sea level pressure & \unit{Pa} \\
10u & 10m wind speed (zonal component) & \unit{m/s} \\
10v & 10m wind speed (meridional component) & \unit{m/s} \\
2t & 2m temperature & \unit{K} \\
2d & 2m dewpoint temperature & \unit{K} \\
src & Surface roughness & \unit{m} \\
skt & Skin temperature & \unit{K} \\
tsn & Top net solar radiation & \unit{W/m^2} \\
\hline
\end{tabular} 
\caption{WRF initialization variables retrieved from the ECMWF subseasonal database. Atmospheric 3D variables (z, u, v, t, q, r) are provided on 10 standard pressure levels: 1000, 925, 850, 700, 500, 300, 200, 100, 50, and 10 hPa. Soil variables (swvl and stl) correspond to the four ECMWF land surface model layers with depths of 0--7 cm (Layer 1), 7--28 cm (Layer 2), 28--100 cm (Layer 3), and 100--289 cm (Layer 4).}
\label{tab:inputs}
\end{table}

\section{The diffusion function}
\label{app:diffusion-function}
In order to gradually erase the original information, the noise level $\sigma(t)$ at any given time $t \in [0, 1]$, is chosen to be a tangent schedule. This approach is based on the cosine schedule introduced by \cite{nichol2021improved}, where scheduling a signal-to-noise ratio term as a cosine function results in the noise level $\sigma(t)$ following a tangent function
\begin{equation}
\sigma(t) = \tan\left(\frac{\pi}{2} t\right), 
\end{equation}
which is scaled to produce noise levels in the range $[0, 100]$.

\section{Sampling procedure}
\label{app:sampling-procedure}

To generate new samples, we solve the backward SDE in (\ref{eq:backward}) using the Euler-Maruyama numerical integrator. The integration is performed over a discretized time span of 256 steps. The noise level schedule follows the exponential decay proposed in \citeA{karras2022elucidating}'s framework, with $\rho=7$. This choice concentrates the discretization steps towards the low-noise regime ($t \approx 0$), allowing the model to spend more computational effort on refining fine details in the sample. The integration proceeds until a minimum noise level of $\sigma_{\text{min}} = 10^{-3}$ is reached, for stability.

\section{Training details}
\label{app:training-details}

Since $\nabla \log p(\boldsymbol{x}_t)$ is generally intractable, it is parameterized as $s_{\boldsymbol{\theta}}(\boldsymbol{x}_t,t)$, where $\boldsymbol{\theta}$ represents the neural network parameters. Following \citeA{vincent2011connection}, the loss function

\begin{equation}
L(\theta,t) = \mathbb{E}_{\boldsymbol{X}_0}\Big[\big|\big|s_\theta(\boldsymbol{X}_t,t) - \nabla \log p(\boldsymbol{X}_t|\boldsymbol{X}_0)\big|\big|_2^2\Big]
\end{equation}
gives an approximation $s_\theta(\boldsymbol{x}_t,t) \approx \nabla \log p(\boldsymbol{x}_t)$. In order to re-write the second term, observe that, using the forward process

\begin{equation}
\boldsymbol{X}_t = \boldsymbol{X}_0 + \sigma(t) \boldsymbol{N}, \qquad \boldsymbol{N} \sim \mathcal{N}(0,I),
\end{equation}
where $\boldsymbol{N}$ represents standard Gaussian noise, and sampling the diffusion time $T \sim \mathcal{U}[0,1]$ uniformly, the practical loss is

\begin{equation}
L(\theta) = \mathbb{E}_{\boldsymbol{X}_0,\boldsymbol{N},T}\Big[\lambda(T)\big|\big|s_\theta(\boldsymbol{X}_0 + \sigma(T) \boldsymbol{N},T) + \tfrac{\boldsymbol{N}}{\sigma(T)}\big|\big|_2^2\Big],
\end{equation}
with $\lambda(t) = 1/\sigma(t)^2$ as a time-dependent weighting function that emphasizes low noise levels. For more information on the architecture of the neural network behind $s_{\boldsymbol{\theta}}(\boldsymbol{x}_t,t)$, see \ref{app:score-architecture}.

During training, noise levels $\sigma$ are sampled from a log-uniform distribution, clipped at a minimum value of $\sigma_{\text{min}}=10^{-4}$. To balance the contribution of different noise levels to the overall loss, we employ the weighting scheme proposed by \citeA{karras2022elucidating}, where the loss at each noise level is weighted by
\begin{equation}
\lambda(\sigma) = \frac{\sigma_{\text{data}}^2 + \sigma^2}{(\sigma \ \sigma_{\text{data}})^2},
\end{equation}
where the key parameter $\sigma_{\text{data}}$ is the standard deviation of the training data, set to $0.31$.

The model is trained for a total of 1,000,000 steps using a batch size of 2. We use the Adam optimizer for parameter updates. The learning rate is managed by a schedule with a linear warm-up followed by a cosine decay. The learning rate starts at 0.0 and increases linearly to a peak value of $10^{-4}$ over the first 1000 warm-up steps. Following this warm-up phase, the learning rate decays following a cosine curve, reaching a final value of $10^{-6}$ at the end of training. To stabilize training and improve the generalization of the final model, we maintain an exponential moving average of the model's parameters with a decay rate of 0.99, and these parameters are used for generating samples.

\section{Score architecture}
\label{app:score-architecture}

The score function is parameterized using a U-Net architecture \cite{ronneberger2015unet}. This network is designed to predict the denoised data from a noised input, and it is explicitly preconditioned by the data's standard deviation $\sigma_\text{data}$.

The U-Net operates on a fixed internal resolution of 224×336, which is chosen to be the closest resolution that is divisible by the number of downsampling layers. The architecture consists of four resolution levels, with channels (32, 64, 128, 256). Each downsampling step reduces the spatial dimensions by a factor of 2. At each resolution level, 6 residual blocks \cite{he2015resnet} are used to process the features.

To incorporate the noise level $\sigma$ at each step, it is first mapped to a 128-dimensional embedding vector using Fourier projections \cite{tancik2020fourier}, which is then passed into the residual blocks. To capture long-range spatial dependencies, multi-head self-attention mechanisms with 8 heads are integrated into the network's bottleneck \cite{vaswani2017attention}.

\clearpage
\section{Supporting figures}
\label{app:supporting-figures}

\begin{figure}
 \centering
 \includegraphics[width=\linewidth]{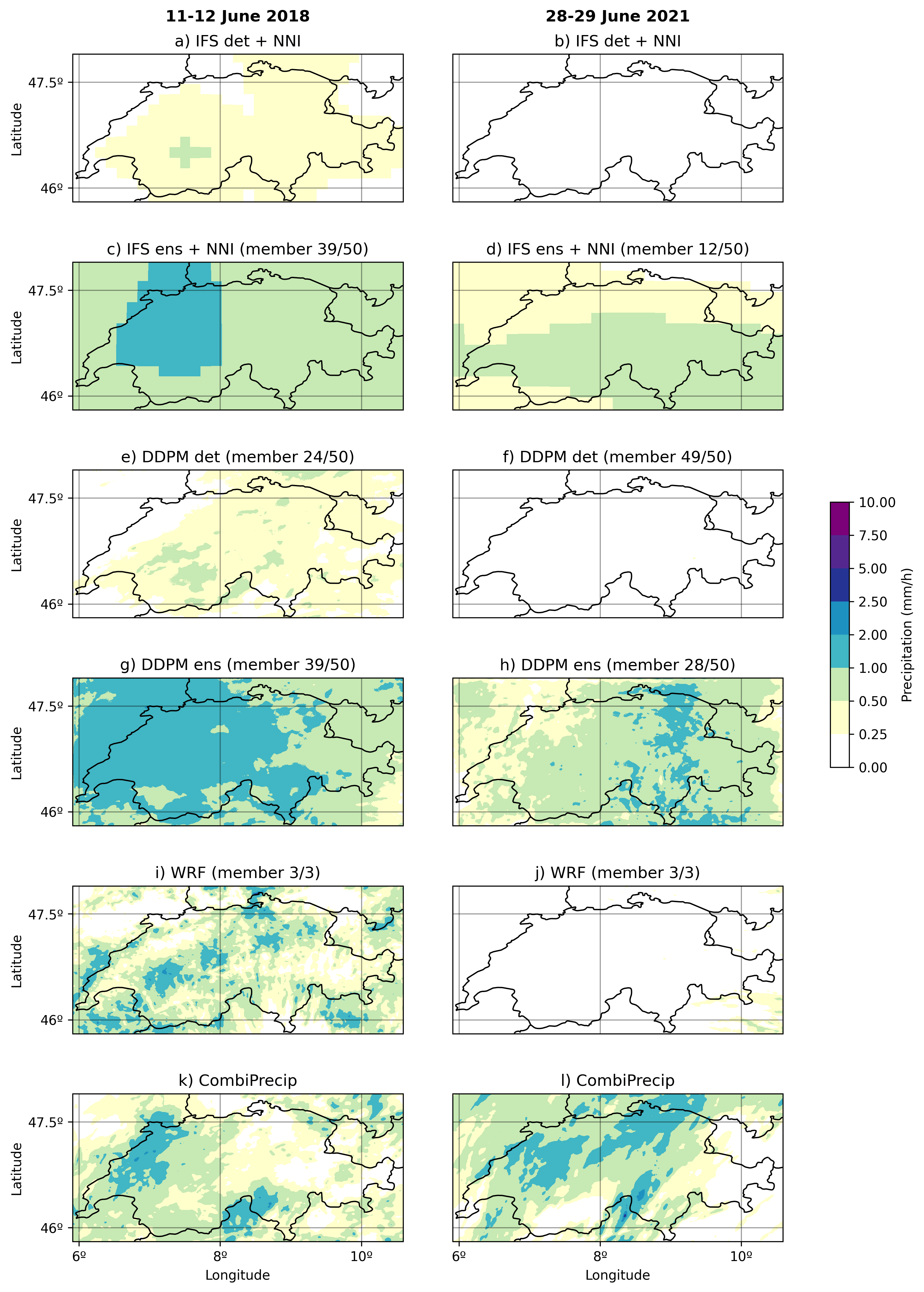}
 \caption{(a-j) Mean aggregated precipitation for all five benchmarked models compared against the (k-l) CombiPrecip ground truth at a 1-week lead time for the 2018 (left column) and 2021 (right column) extreme precipitation events. Rows from top to bottom display (a-b) IFS det + NNI, (c-d) IFS ens + NNI, (e-f) DDPM det, (g-h) DDPM ens, (i-j) WRF. For ensemble models, the displayed member corresponds to the one with the highest total accumulated precipitation over the event window.}
 \label{figs:maps-1-week}
\end{figure}

\begin{figure}
 \centering
 \includegraphics[width=\linewidth]{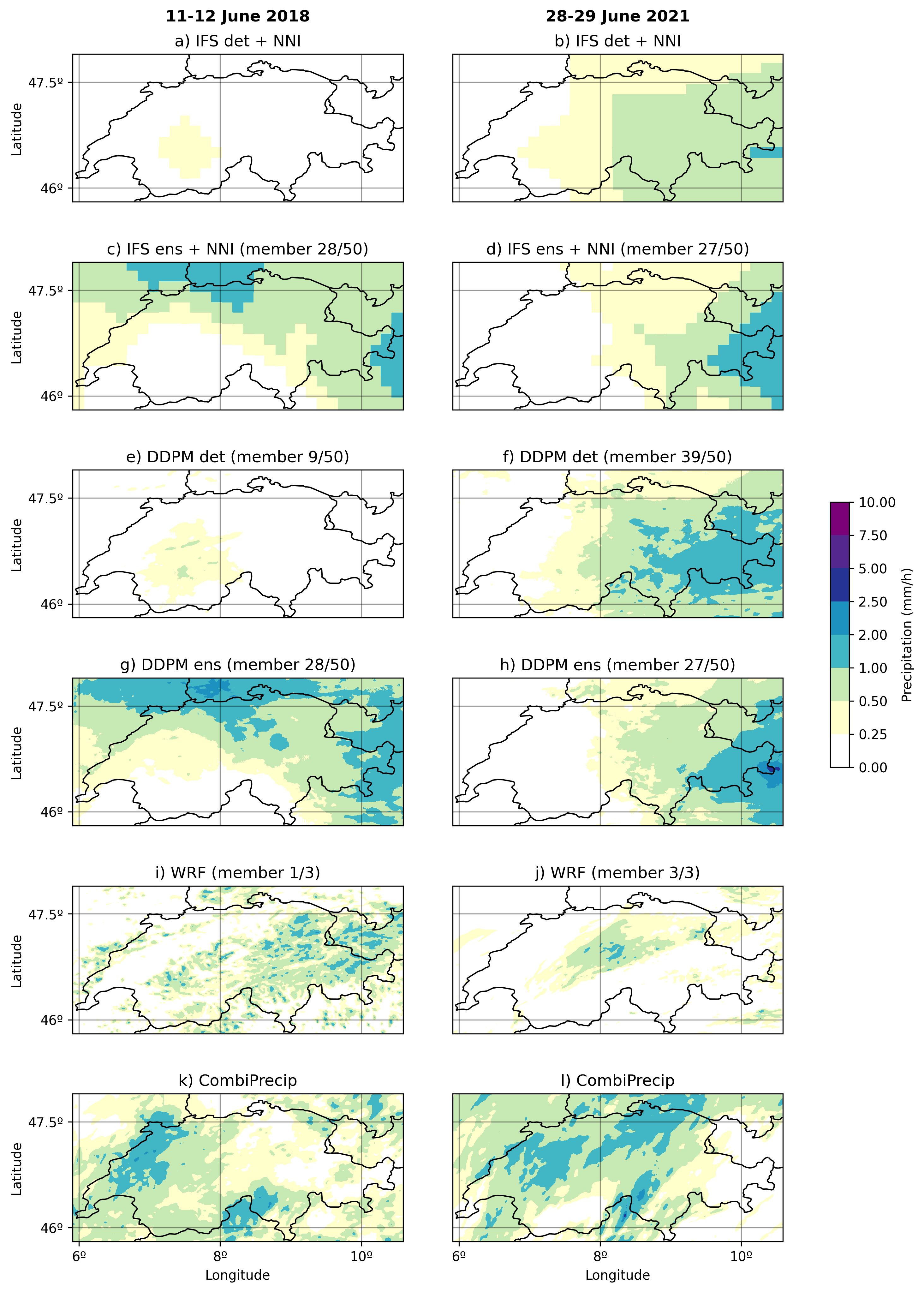}
 \caption{(a-j) Mean aggregated precipitation for all five benchmarked models compared against the (k-l) CombiPrecip ground truth at a 2-week lead time for the 2018 (left column) and 2021 (right column) extreme precipitation events. Rows from top to bottom display (a-b) IFS det + NNI, (c-d) IFS ens + NNI, (e-f) DDPM det, (g-h) DDPM ens, (i-j) WRF. For ensemble models, the displayed member corresponds to the one with the highest total accumulated precipitation over the event window.}
 \label{figs:maps-2-week}
\end{figure}

\begin{figure}
 \centering
 \includegraphics[width=\linewidth]{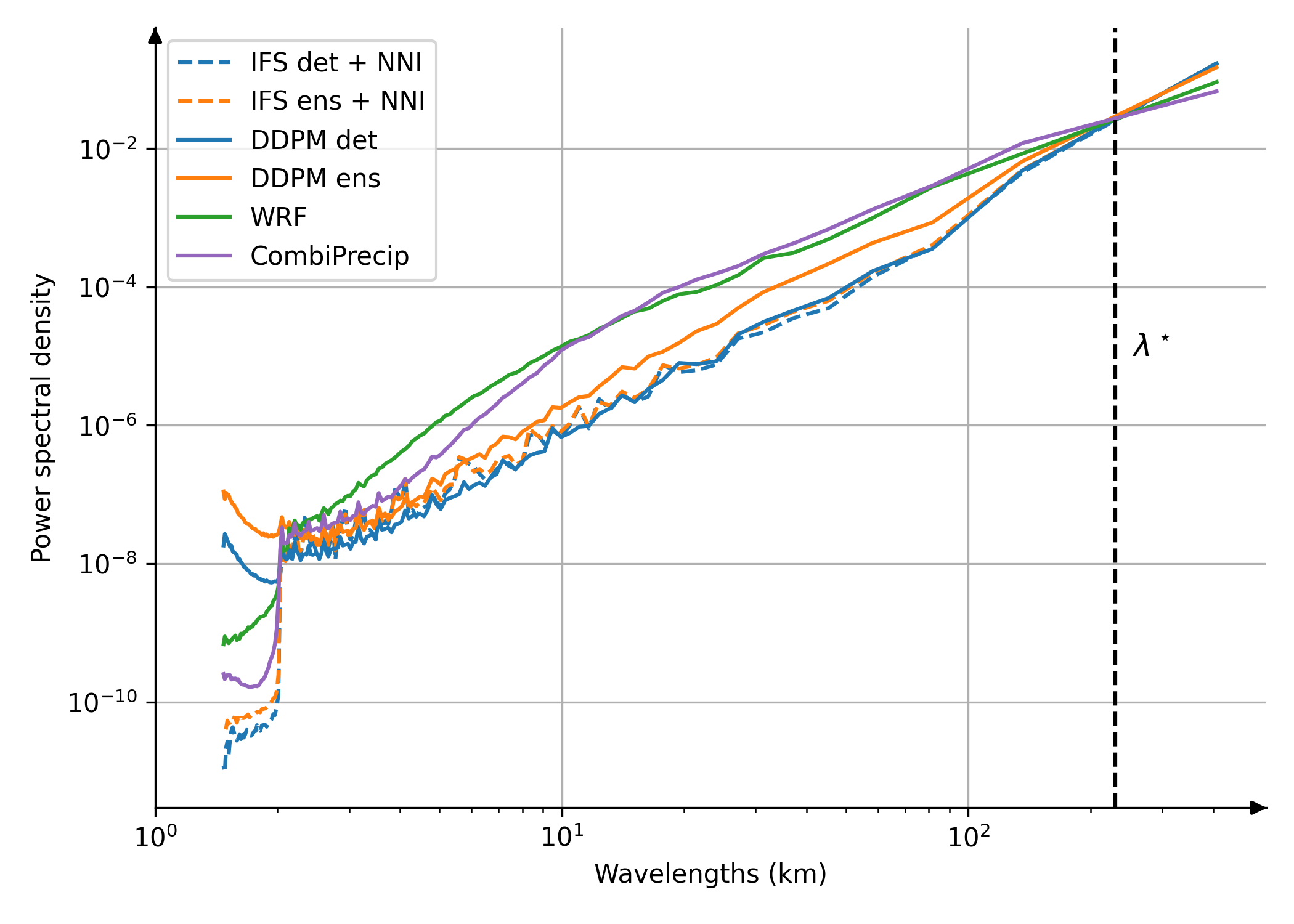}
 \caption{Power spectral density of the 5 benchmarked models in this work for the 1-week lead time as well as the ground truth provided by CombiPrecip. Here, the two events are concatenated over the time axis.}
 \label{figs:psd-1-week}
\end{figure}

\begin{figure}
 \centering
 \includegraphics[width=\linewidth]{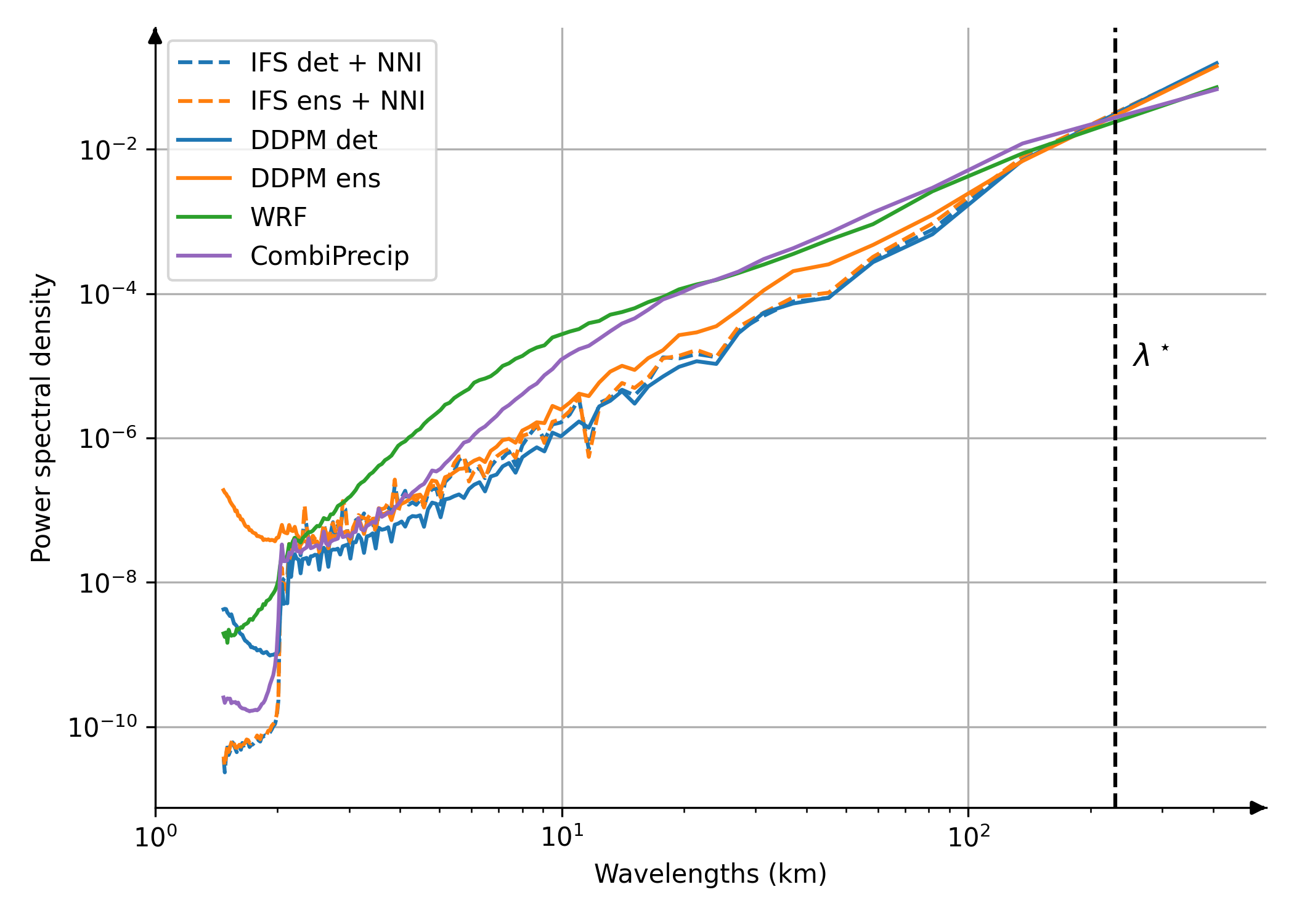}
 \caption{Power spectral density of the 5 benchmarked models in this work for the 2-week lead time as well as the ground truth provided by CombiPrecip. Here, the two events are concatenated over the time axis.}
 \label{figs:psd-2-week}
\end{figure}

\end{document}